\documentclass [11pt]{article}
 \pdfoutput=1
\usepackage{amsfonts}
\usepackage{graphicx}
\usepackage{amsmath,bm}
\usepackage{amssymb}
\usepackage{latexsym}
\usepackage{mathrsfs}
\usepackage{color}
\usepackage{slashed}
\definecolor{red}{rgb}{1,0,0}
\usepackage{cancel}
\usepackage[nosort]{cite}
\usepackage{color}

\usepackage{soul} 

\usepackage{setspace} 

\definecolor{red}{rgb}{1,0,0}

\makeatletter
\def\section{\@startsection {section}{1}{\z@}{-3.5ex plus -1ex minus
 -.2ex}{2.3ex plus .2ex}{\large\bf}}
\def\subsection{\@startsection{subsection}{2}{\z@}{-3.25ex plus -1ex
minus -.2ex}{1.5ex plus .2ex}{\normalsize\bf}}
\makeatother
\makeatletter

\@addtoreset{equation}{section}

\makeatother

\def\bel{\begin{equation}\begin{aligned}}
\def\eel{\end{aligned}\end{equation}}

\def\bea{\begin{eqnarray}} \def\eea{\end{eqnarray}}
\def\be{\begin{equation}} \def\ee{\end{equation}} 
\def\nn{\nonumber}

\newcommand{\promille}{%
  \relax\ifmmode\promillezeichen
        \else\leavevmode\(\mathsurround=0pt\promillezeichen\)\fi}
\newcommand{\promillezeichen}{%
  \kern-.05em%
  \raise.5ex\hbox{\the\scriptfont0 0}%
  \kern-.15em/\kern-.15em%
  \lower.25ex\hbox{\the\scriptfont0 00}}

\usepackage[colorlinks,linkcolor=black,citecolor=blue,urlcolor=blue,linktocpage]{hyperref}

\begin{document}
\setcounter{page}{0}
\thispagestyle{empty}

\parskip 3pt

\font\mini=cmr10 at 2pt

\begin{titlepage}
~\vspace{1cm}
\begin{center}

{\LARGE \bf The Power of Perturbation Theory}

\vspace{0.6cm}

{\large
Marco~Serone$^{a,b}$, Gabriele~Spada$^a$, and Giovanni~Villadoro$^{b}$}
\\
\vspace{.6cm}
{\normalsize { \sl $^{a}$ 
SISSA International School for Advanced Studies and INFN Trieste, \\
Via Bonomea 265, 34136, Trieste, Italy }}

\vspace{.3cm}
{\normalsize { \sl $^{b}$ Abdus Salam International Centre for Theoretical Physics, \\
Strada Costiera 11, 34151, Trieste, Italy}}

\end{center}
\vspace{.8cm}
\begin{abstract}
We study quantum mechanical systems with a discrete spectrum.
We  show that the asymptotic series associated to certain paths of steepest-descent
(Lefschetz thimbles) are Borel resummable to the full result.
Using a geometrical approach based on the Picard-Lefschetz theory
we characterize the conditions under which perturbative expansions lead to exact results.
Even when such conditions are not met, 
we explain how to define a different perturbative expansion that 
reproduces the full answer without the need of transseries, 
i.e. non-perturbative effects, such as real (or complex) instantons.
Applications to several quantum mechanical systems are presented.
\end{abstract}

\end{titlepage}

\tableofcontents

\section{Introduction}

It is known since ref.\cite{Dyson:1952tj} that perturbative expansions in quantum field theories (QFT), as well as in quantum mechanics (QM), 
are generically asymptotic with zero radius of convergence. In special cases, such as the anharmonic oscillator in QM and $\phi^4$ theories up to $d=3$ space-time dimensions, the perturbative expansion
turns out to be Borel resummable \cite{Loeffel:1970fe,Simon:1970mc,Eckmann,Magnen}. For the anharmonic oscillator it has been verified that  
the Borel resummed perturbative series converges to the exact result, 
available by other numerical methods.
Perturbative series associated to more general systems and/or in higher space-time dimensions are typically non-Borel resummable, because of singularities in
the domain of integration. These can be avoided by deforming the contour 
at the cost of introducing an ambiguity that is  non-perturbative in the expansion parameter $\lambda$. The ambiguity is expected to be removed by including
contributions from semiclassical instanton-like configurations 
(and all their corresponding series expansion), resulting in what is called transseries.
There has been considerable progress in recent years on these issues in the context of 
the theory of resurgence  \cite{ecalle} (see e.g. ref.\cite{Aniceto:2011nu}, and refs.\cite{Marino:2012zq,Dorigoni:2014hea} for reviews and further references).
A systematic implementation to generic QFT and QM is however not straightforward.
A resurgent analysis requires a detailed knowledge of the asymptotic form of the perturbative coefficients,
while typically only the leading large-order behaviour of the perturbative expansion might be accessed in generic QFT and QM \cite{Lipatov:1976ny,Brezin:1976vw,Brezin:1976wa,Brezin:1977gk}. 
Besides, the knowledge of the coefficients of the perturbative series alone
is not enough to guarantee that the reconstructed transseries reproduces the full answer. 
Some non-perturbative information is required, such as the knowledge of some analytic properties 
of the observable as a function of the expansion parameter.
Most importantly, the practicality of transseries beyond the weak coupling regime 
is hindered by the need to resum the series expansion 
of all the semi-classical configurations that contribute, in general infinite in number.

Perturbation theory within a path integral formulation is an infinite dimensional generalization of the usual steepest-descent method to evaluate ordinary integrals. For sufficiently regular functions 
Picard-Lefschetz theory teaches us how to decompose the initial contour of integration 
into a sum of steepest-descent trajectories (called Lefschetz thimbles, or simply thimbles).
A geometric approach to the path integral from this perspective, as well as an excellent introduction for physicists to these ideas, has been given by Witten \cite{Witten:2010cx}
(see also refs.\cite{Witten:2010zr,Harlow:2011ny}).
The theory of Lefschetz thimbles allows us to rigorously classify which saddle-point configurations contribute to a given physical observable.

An interesting question to ask is under what conditions no non-trivial saddle point contributes, so that the whole result is given by the single perturbative series around  the trivial saddle-point. In terms of Lefschetz thimbles, this corresponds to the simple situation in which the domain of integration of
the path integral  does not need any deformation being already a single Lefschetz thimble on its own. This is what should happen for instance in the anharmonic oscillator in which, as we mentioned, the perturbative series is Borel resummable and converges to the exact result.

It has recently been shown in ref.\cite{shortpaper} that several one-dimensional quantum mechanical models with a discrete spectrum admit an ``exact perturbation theory'' (EPT) that is able to capture the full result including non-perturbative effects, even in cases which are known to receive instanton corrections, such as the (supersymmetric) double well. 

In this paper we explain the reasons behind the results of ref.\cite{shortpaper}, 
using the path integral formulation and a Lefschetz thimble perspective.
For pedagogical purposes, in sec.~\ref{sec:ODI}  we start by reviewing 
the concepts of Borel summability
and Lefschetz-thimble decomposition for a class of 
one-dimensional integrals $Z(\lambda)$, viewed as 0-dimensional path integrals,
with the parameter $\lambda$ playing the role of $\hbar$. 
In fact, all the properties of perturbation theory, the role of non-perturbative saddles
as well as the definition of EPT can easily be understood in this context. 
The Lefschetz thimble decomposition reduces $Z(\lambda)$ into a sum
of integrals over thimbles---steepest descent paths with
a single saddle point. We prove that their saddle-point expansion 
is always Borel resummable to the exact answer.
In contrast to previous works in the literature, there is no need to study the analytic properties of the integral 
as a function of $\lambda$. Indeed,  thanks to a suitable change of coordinates, we are able to rewrite the integral over thimbles directly in terms
of a well-defined Borel transform. 
This result implies the following important consequences.
When the decomposition of $Z(\lambda)$ involves trivially only one thimble,
its ordinary perturbation theory is also Borel resummable to the whole result.
On the contrary, when the decomposition involves more than one thimble, 
or it requires an analytic continuation in $\lambda$, 
the naive series expansion of $Z(\lambda)$ 
is not Borel resummable to the exact answer.

Independently of the thimble decomposition of $Z(\lambda)$, 
we show how to introduce a second integral $\hat Z(\lambda,\lambda_0)$
which has a trivial thimble decomposition for any fixed $\lambda_0$ 
and coincides with $Z(\lambda)$ at $\lambda_0=\lambda$. 
Therefore the expansion of $\hat Z(\lambda,\lambda_0)$ in $\lambda$ is Borel
resummable to the exact result even when this is not the case for $Z(\lambda)$.
Such result is possible considering that $Z(\lambda)$ and $\hat Z(\lambda,\lambda_0)$, at fixed $\lambda_0$,  have different analytical properties in $\lambda$.
The expansion of $\hat Z(\lambda,\lambda_0)$  is the simplest implementation of EPT.

In sec.~\ref{sec:PI} the Borel summability of thimbles is readily extended to multi-dimensional 
integrals and we discuss in some detail the non trivial generalization to path integrals in QM. In this way we are able to show that QM systems with a bound-state potential and a single non-degenerate crtitical point---the anharmonic oscillator being the prototypical example---are entirely reconstructable from their perturbative expansion. 
Namely, for any observable (energy eigenvalues, eigenfunctions, etc.)
the asymptotic perturbation theory is Borel resummable to the exact 
result.\footnote{As far as we know, the Borel resummability of 
observables other than the energy levels has not received much 
attention in the literature.} 
At least for the ground state energy, this remains true also for potentials
with multiple critical points as long as the absolute minimum is unique.

Potentials $V(x;\lambda)$ with more than one critical point are more problematic because 
not all observables are Borel resummable to the exact result and 
in general instantons are well-known to contribute.
Unfortunately in most situations it is a challenging task to explicitly classify all saddle-points and evaluate the corresponding contributions (see e.g. ref.~\cite{Tanizaki:2014xba} for a recent attempt). In analogy to the 
one-dimensional integral we show how to bypass this problem by considering
an alternative potential $\hat V(x;\lambda,\lambda_0)$ admitting always 
a Borel resummable perturbation theory in $\lambda$ and coinciding 
to the original one for $\lambda_0=\lambda$. The idea is to choose $\hat V$
as the sum of a tree-level and a quantum potential, with the former having
only a single critical point. Since the thimble decomposition is controlled only
by the saddle point of the tree-level part, the perturbative expansion of 
$\hat V$ (EPT) is guaranteed to be Borel resummable to the exact result. 

For any value of the coupling constant $\lambda$, EPT captures the full result. 
In contrast, the expansion from $V$ requires in general  also the inclusion 
of instanton contributions,  we denote such expansion 
Standard Perturbation Theory (SPT).
As noticed also in ref.~\cite{shortpaper},  EPT works surprisingly well at strong coupling, where SPT becomes impractical.

In the spirit of resurgence the coefficients of the perturbative series encode the exact answer, with the crucial difference that no transseries are needed.
Using this method, we can relax the requirement of having a single critical point in the original potential
$V$, and arrive to the following statement:
{\it 
In one-dimensional QM systems with a bound-state potential $V$ 
 that admits the $\hat V$ defined above, any observable can be 
 exactly computed from a single perturbative series.} 

We illustrate our results in sec.~\ref{sec:examples} by a numerical study of the following quantum mechanical examples: the (tilted) anharmonic potential, the symmetric double well, its supersymmetric version, the perturbative expansion around a false vacuum, and pure anharmonic oscillators.\footnote{After the completion of this work we became aware of ref.\cite{Caswell:1979qh} where manipulations similar to the ones done in this paper to define EPT have been proposed in the context of anharmonic oscillators and symmetric double-well potentials.  
However, the generality of the approach and the conditions ensuring Borel summability to the exact result of the perturbative series have not been spelled out in ref.\cite{Caswell:1979qh}.}
 
In all these systems  we will show that the exact ground state energy, computed by solving the Schrodinger equation, is recovered without the need of advocating 
non-perturbative effects, such as real (or complex) instantons.
We will also show that the same applies for higher energy levels and the eigenfunctions.

We conclude in sec.~\ref{sec:concl}, where we also briefly report the future perspectives to extend our results in QFT.
Some technical details associated to sec.~\ref{sec:PI} are reported in an appendix.

\section{One-Dimensional Integrals}

\label{sec:ODI}

The main points of this paper are best understood by considering one-dimensional integrals, where a step-by-step analytical study is possible. 
In order to be self-contained, we first review essential facts about Lefschetz thimbles and Borel resummation methods  in subsecs. \ref{subsec:LTD} and \ref{subsec:ASBS}, respectively. 
Readers familiar with these topics might jump directly to subsec. \ref{subsec:ITBR}.

\subsection{Lefschetz-Thimble Decomposition}

\label{subsec:LTD}

Consider the integral of the type
\begin{equation} 
Z(\lambda) \equiv \frac{1}{\sqrt{\lambda}}\int_{-\infty}^{\infty} dx\, g(x)\, e^{-f(x)/\lambda}\,,
\label{IgGen}
\end{equation}
one-dimensional prototype of path-integrals in QM and QFT.
We assume that the functions $g(x)$ and $f(x)$, in general complex, are regular
and the convergence of the integral for positive values of $\lambda$ is determined only by 
$f(x)$.\footnote{We assume this to be true also for the analytic continuation
of the integrand on the complex $x$-plane, which we will perform soon.}
In general $g$ might also present a sufficiently regular dependence on $\lambda$.
For simplicity, we take $f$ and $g$ to be entire functions of $x$, though more general cases could be considered. 

The perturbative expansion of $Z(\lambda)$ around $\lambda=0$ corresponds to the saddle-point
approximation of the integral (\ref{IgGen}).\footnote{Note that if $g(x)$ is brought to the exponent
the saddle points of $f(x)-\lambda \log g(x)$ will be different. The associated saddle-point expansion, however, will not correspond to the original expansion in $\lambda$. } 
Since the function $f$ in general has multiple saddle points
and each saddle point has its own perturbative expansion, the exact result for $Z(\lambda)$ is recovered
by a non-trivial combination of the various saddle-point contributions (properly resummed).
We will review in this subsection the theory describing how to combine the various saddle-points, and 
discuss in the next one how to exactly resum each expansion. 

The idea is to deform the integration contour into a sum of steepest descent paths of the saddle points.
As first step we analytically continue the functions $f$ and $g$ in the complex plane $z=x+i y$
and view eq.~(\ref{IgGen}) as an open contour integral in $z$: 
\be
Z(\lambda) =\frac{1}{\sqrt{\lambda}} \int_{{\cal C}_x} \!\! dz\, g(z)\,  e^{-f(z)/\lambda}, 
\label{IgDef}
\ee
where ${\cal C}_x$ is the real axis. We call $z_\sigma$ the saddle points of $f(z)$, i.e $f^\prime(z_\sigma)=0$.
As long as $z_\sigma$ are isolated and non-degenerate, $f^{\prime\prime}(z_\sigma)\neq 0$,  the contour of steepest-descent passing through $z_\sigma$ is determined by a flow 
line $z(u)$ satisfying the first-order equations
\be
\frac{d z}{du} = \eta \frac{\partial \overline F}{\partial \bar z}\,,  \ \ \ \ \frac{d \bar z}{du} = \eta \frac{\partial F}{\partial z}\,, \ \ \eta=\pm 1\,,
\label{udfe}
\ee
where $F(z)\equiv -f(z)/\lambda$ and  $u$ is the real line parameter. 
Unless $z(u)=z_\sigma$ for all $u$, a non-constant flow can reach $z_\sigma$ only
for $u=\pm \infty$. Using eq.~(\ref{udfe}) one has
\be
\frac{d F}{du} = \frac{\partial F}{\partial z} \frac{dz}{du} = \eta  \Big| \frac{\partial F}{\partial z}\Big|^2 \,.
\label{udfe2}
\ee
The cycles with $\eta = -1$ and $\eta = +1$ are denoted respectively downward and upward flows, since  ${\rm Re}\, F$ is monotonically decreasing and increasing in the two cases, 
as eq.~(\ref{udfe2}) indicates. 
Following the notation of ref.\cite{Witten:2010cx}\footnote{We refer the reader to sec. 3 of this paper for a more extensive introduction to Lefschetz thimbles.} we denote by ${\cal J}_\sigma$ and  ${\cal K}_\sigma$ the downward and upward flows passing through the 
saddle point $z_\sigma$.
Equation (\ref{udfe2}) shows  that ${\rm Im}\,F$ is constant on both ${\cal J}_\sigma$ and  ${\cal K}_\sigma$.
The downward flow ${\cal J}_\sigma$ coincides with the path of steepest-descent and
when such path flows to Re~$F=-\infty$ it is called Lefschetz thimble, or thimble for short.
By construction the integral over each thimble is well defined and convergent.
When instead the steepest descent path hits another saddle point, the flow splits into
two branches and an ambiguity arises. The corresponding integral is said to be 
on a Stokes line and, as we will see below, some care is required.

Given the absence of singularities on the complex plane, the contour ${\cal C}_x$
can be freely deformed to match a combination ${\cal C}$ of steepest descent paths keeping the integral (\ref{IgDef}) finite during the deformation:
\be
{\cal C} = \sum_\sigma {\cal J}_\sigma n_\sigma\,. \qquad 
\label{Cdecomp}
\ee
By means of the Picard-Lefschetz theory the integer coefficients $n_\sigma$ are given by 
\be
n_\sigma = \langle {\cal C}_x, {\cal K}_\sigma \rangle  \,,
\label{nsigma}
\ee
where $ \langle {\cal C}_x, {\cal K}_\sigma \rangle$ denote the intersection pairings between the original contour ${\cal C}_x$ and the upward flows ${\cal K}_\sigma$
and we used the fact that ${\cal J}_\sigma$ and  ${\cal K}_\sigma$ are 
dual to each other:
\be
\langle {\cal J}_\sigma, {\cal K}_\tau \rangle = \delta_{\sigma \tau}\,.
\label{DualPair}
\ee
The original integral (\ref{IgGen}) is then reduced to a sum of integrals along 
the thimbles ${\cal J}_\sigma$,
\be
Z(\lambda)=\sum_\sigma n_\sigma Z_\sigma(\lambda)\,, 
\label{IgDefTh}
\ee
where
\be
Z_\sigma(\lambda) \equiv \frac{1}{\sqrt{\lambda}} 
\int_{{\cal J}_\sigma} \!\! dz\, g(z)\,  e^{-f(z)/\lambda}\,.
\label{IgDefTh2}
\ee
Contrary to the naive expectation that the contour of integration should be deformed 
to pass through {\it all} (complex and real) saddles of $f$, only the subset of saddles
with $n_\sigma\neq 0$ must be considered.

In the presence of a flow connecting two saddle points 
$z_{\sigma}$ and $z_\tau$,  we have ${\cal J}_\sigma= {\cal K}_\tau$  and
the corresponding intersection $\langle {\cal J}_\sigma, {\cal K}_\tau \rangle$ 
is not well defined. This problem can be avoided by taking $\lambda$ to be complex, modifying in this way the flow curves, that 
implicitly depend on $\lambda$.
The initial integral is then recovered in the limit Im~$\lambda\to0$.
When $Z(\lambda)$ is not on a Stokes line the intersection numbers $n_\sigma$ in eq.~(\ref{Cdecomp}) are unambiguous in such limit. On a Stokes line instead some of the
$n_\sigma$ are discontinuous and the decomposition (\ref{Cdecomp}) is different in the two limits Im~$\lambda\to0^{\pm}$, yet the same $Z(\lambda)$ is recovered in the two cases.

Two choices of $f$ are particularly interesting for the 
discussion of path integrals in QM and QFT:
$f$ purely imaginary (corresponding to the real-time path integral) and $f$ real 
(corresponding to the Euclidean path integral). In the first case 
the integration cycle ${\cal C}_x$ is not a Lefschetz thimble (the imaginary part is not constant) and the decomposition (\ref{Cdecomp}) is non-trivial.
On the contrary, in the second case $f$ has at least one real saddle and ${\cal C}_x$ coincides with one or more steepest descent paths (being ${\rm Im}\, F=0$). If the real saddle is unique, all others being complex, the real axis is a thimble and ${\cal C}={\cal C}_x$. In presence of more real saddles $Z(\lambda)$ is on a Stokes line 
and the decomposition  (\ref{Cdecomp}) requires an analytic continuation.

The quantum mechanical path integral generalization of this result implies an important difference between Minkoswki and Euclidean times.
While in the former we expect in general  a very complicated Lefschetz thimble decomposition (\ref{Cdecomp}) with an infinite number of saddles contributing, in the latter there is a class of theories where the original integration domain is already a thimble and eq.~(\ref{Cdecomp}) is not necessary.
For this reason we will focus on real functions $f$ and correspondingly we will consider euclidean path integrals.

It is useful to illustrate the discussion above by considering the explicit example
of the integral (\ref{IgGen}) with
\be
f(x,m) =  \frac 12 m \, x^2 + \frac 14 x^4\,, \quad \quad g(x)= 1\,,
\label{Ex1}
\ee
which corresponds to the zero-dimensional reduction of the anharmonic oscillator for $m> 0$, 
the pure anharmonic oscillator for $m=0$ and the symmetric double well for $m<0$. The resulting function $Z(\lambda,m)$ is analytic in $m$ and 
can be written as 
\be
Z(\lambda,m) = \left \{ \begin{array}{lc} 
\sqrt{\frac{m}{2\lambda}} e^{\frac{m^2}{8\lambda}} K_{\frac 14}\left(\frac{m^2}{8\lambda}\right) & m>0\,, \vspace{10pt} \\ 
\frac{\Gamma\left( 1/4 \right ) }{\sqrt2}\, \lambda^{-1/4} & m=0 \,,
\vspace{8pt} \\
 \sqrt{\frac{-m\pi^2}{4\lambda}} e^{\frac{m^2}{8\lambda}} \left [ I_{-\frac 14}\left(\frac{m^2}{8\lambda}\right)+I_{\frac 14}\left(\frac{m^2}{8\lambda}\right)\right ] \quad  & m<0 \,,
\end{array} \right. 
\label{IgEx}
\ee
where $K_n$ and $I_n$ are the modified Bessel functions. 

\begin{figure}[t!]
\begin{center}
\begin{minipage}{0.38\linewidth}
\begin{center}
              \includegraphics[width=55mm]{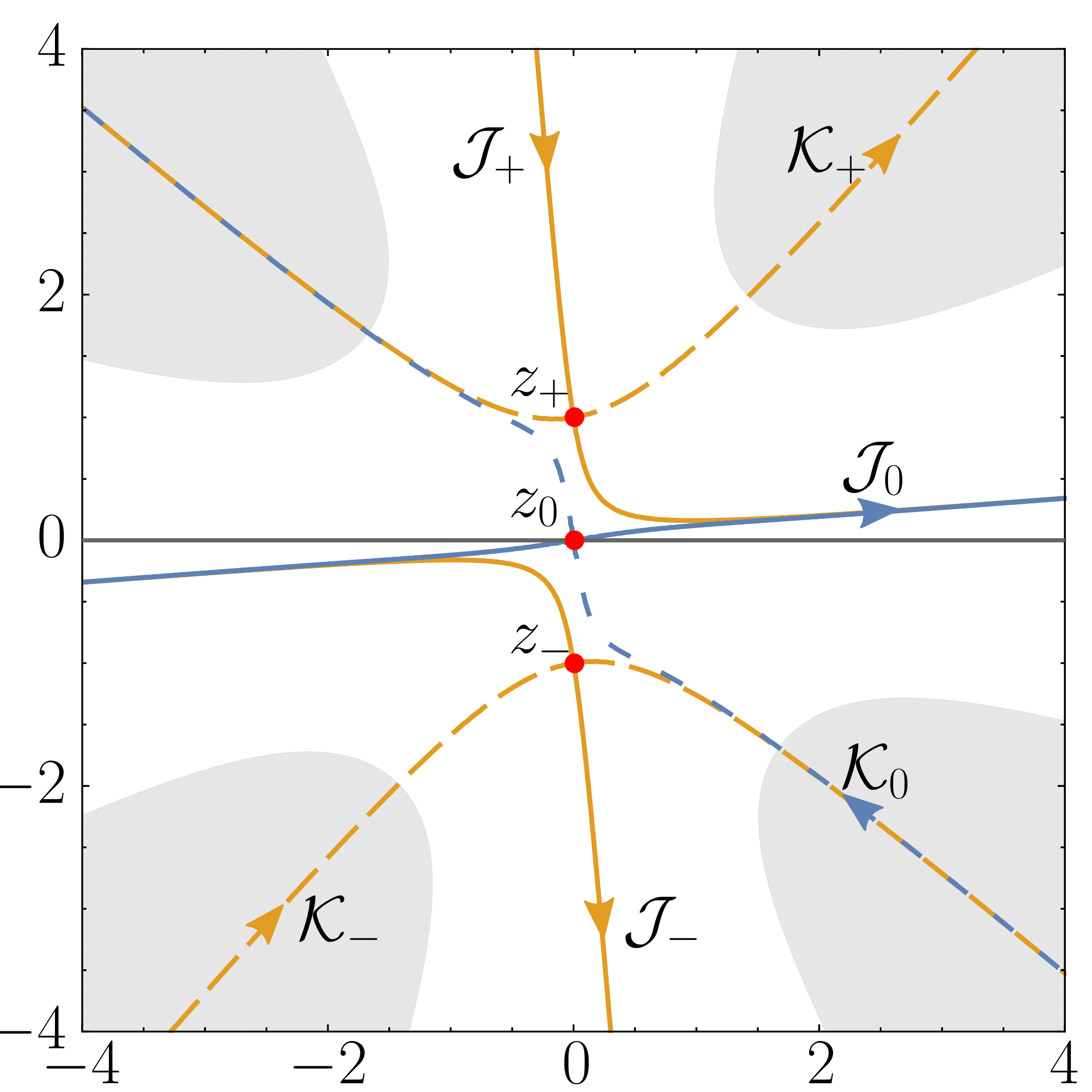}\\
\end{center}
\end{minipage}
%
\begin{minipage}{0.40\linewidth}
\begin{center}
           \includegraphics[width=57mm]{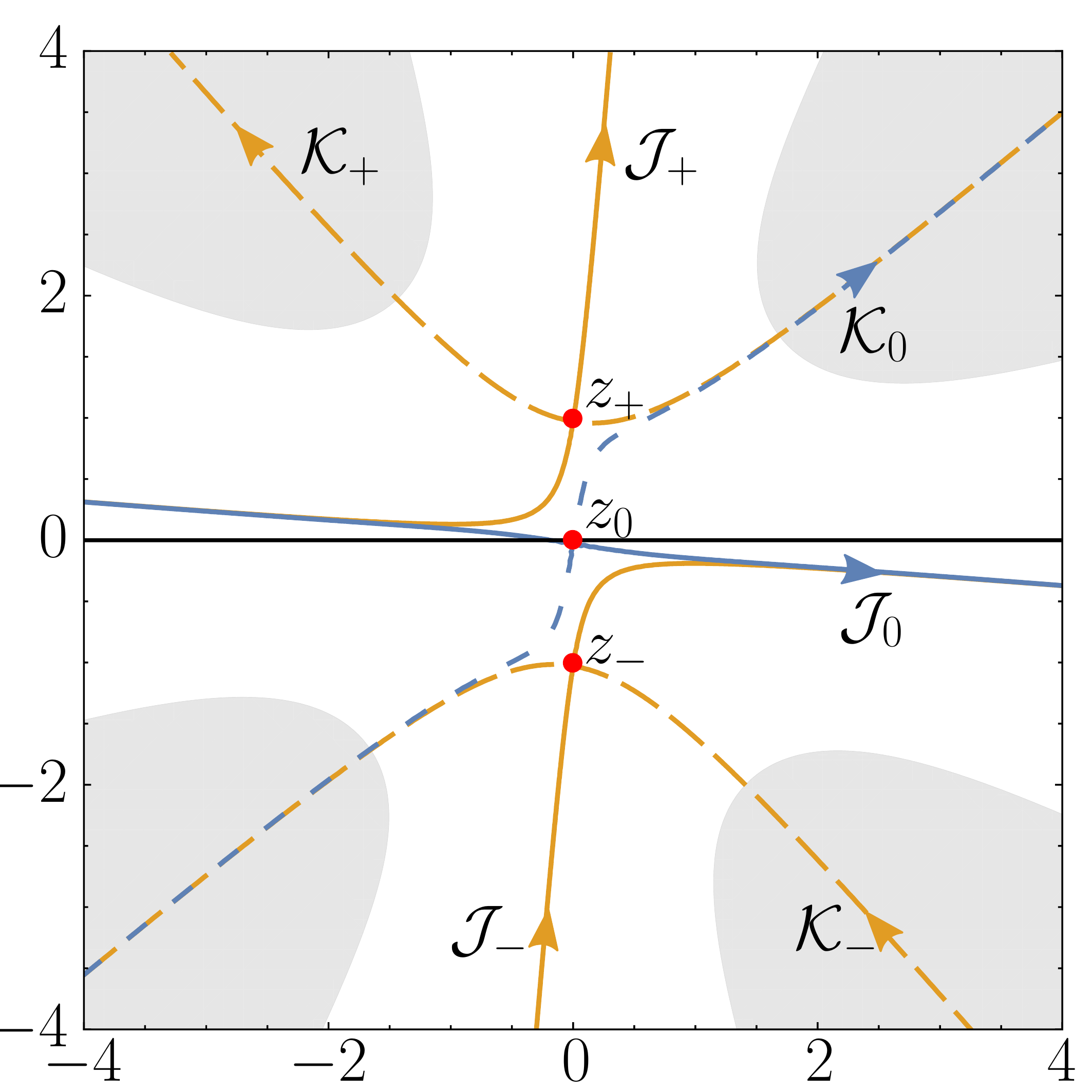}\\
\end{center}
\end{minipage}
\vskip 10pt
\begin{minipage}{0.4\linewidth}
\begin{center}
              \includegraphics[width=57mm]{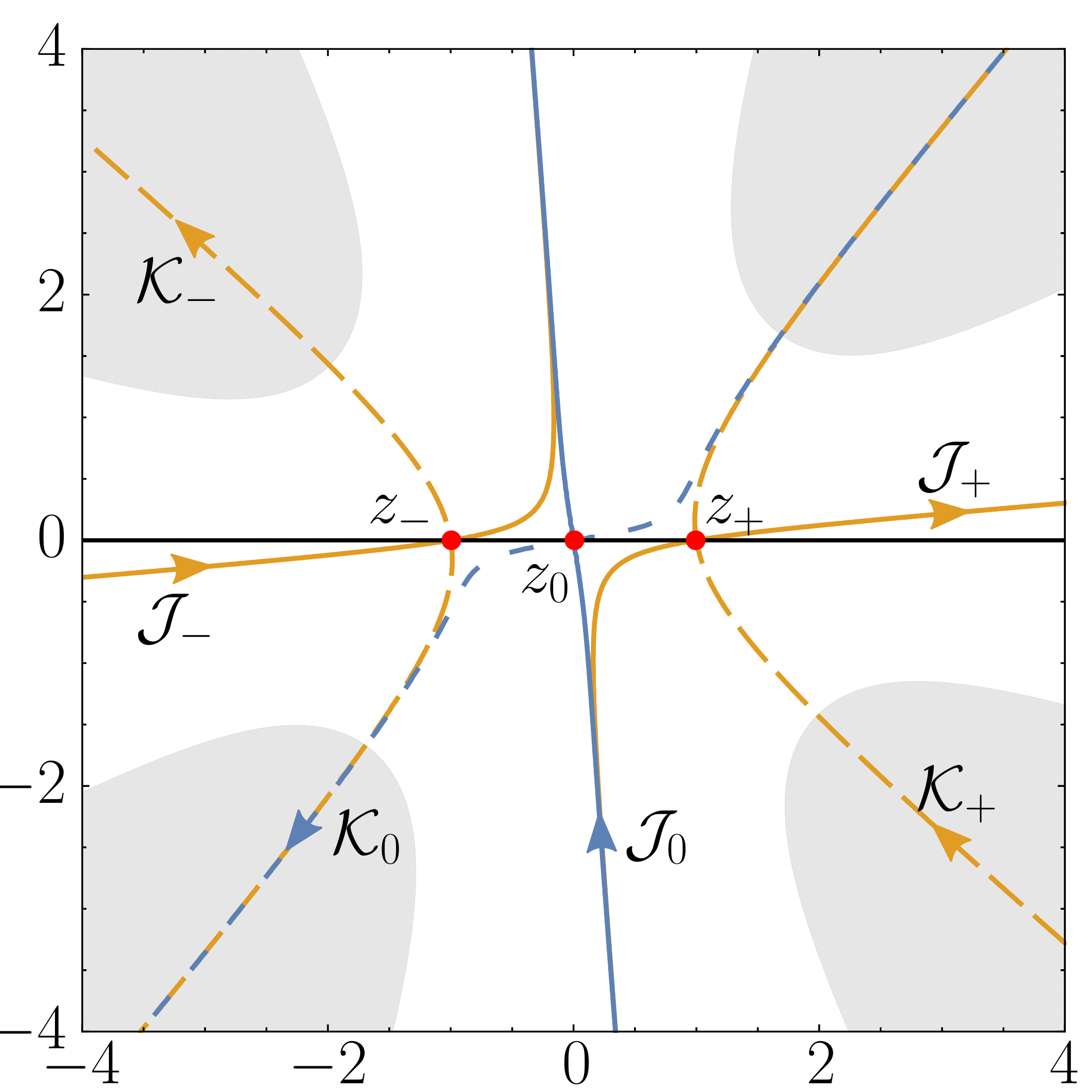}\\
\end{center}
\end{minipage}
%
\begin{minipage}{0.4\linewidth}
\begin{center}
           \includegraphics[width=57mm]{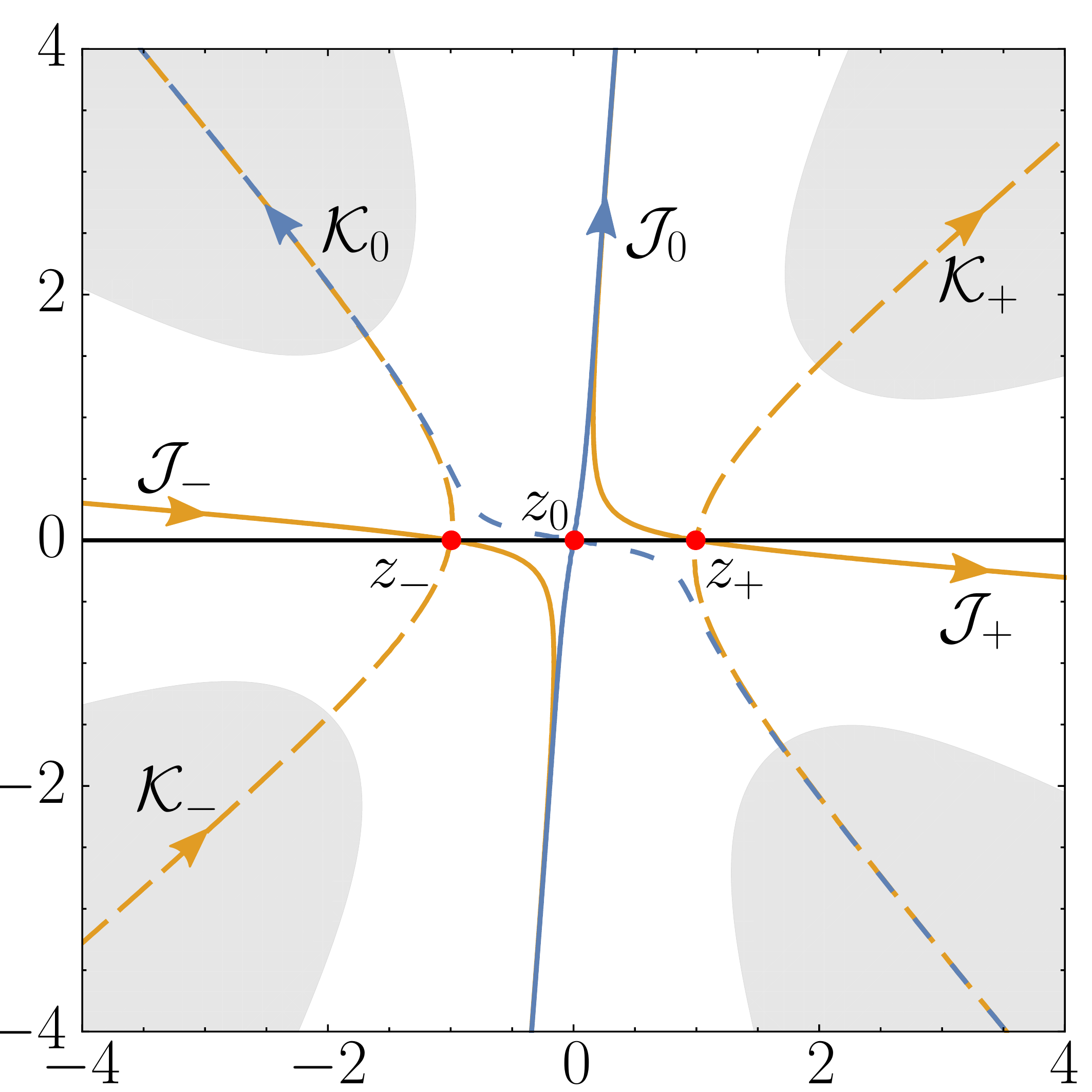}\\
\end{center}
\end{minipage}
\vskip 10pt
\begin{minipage}{0.4\linewidth}
\begin{center}
              \includegraphics[width=57mm]{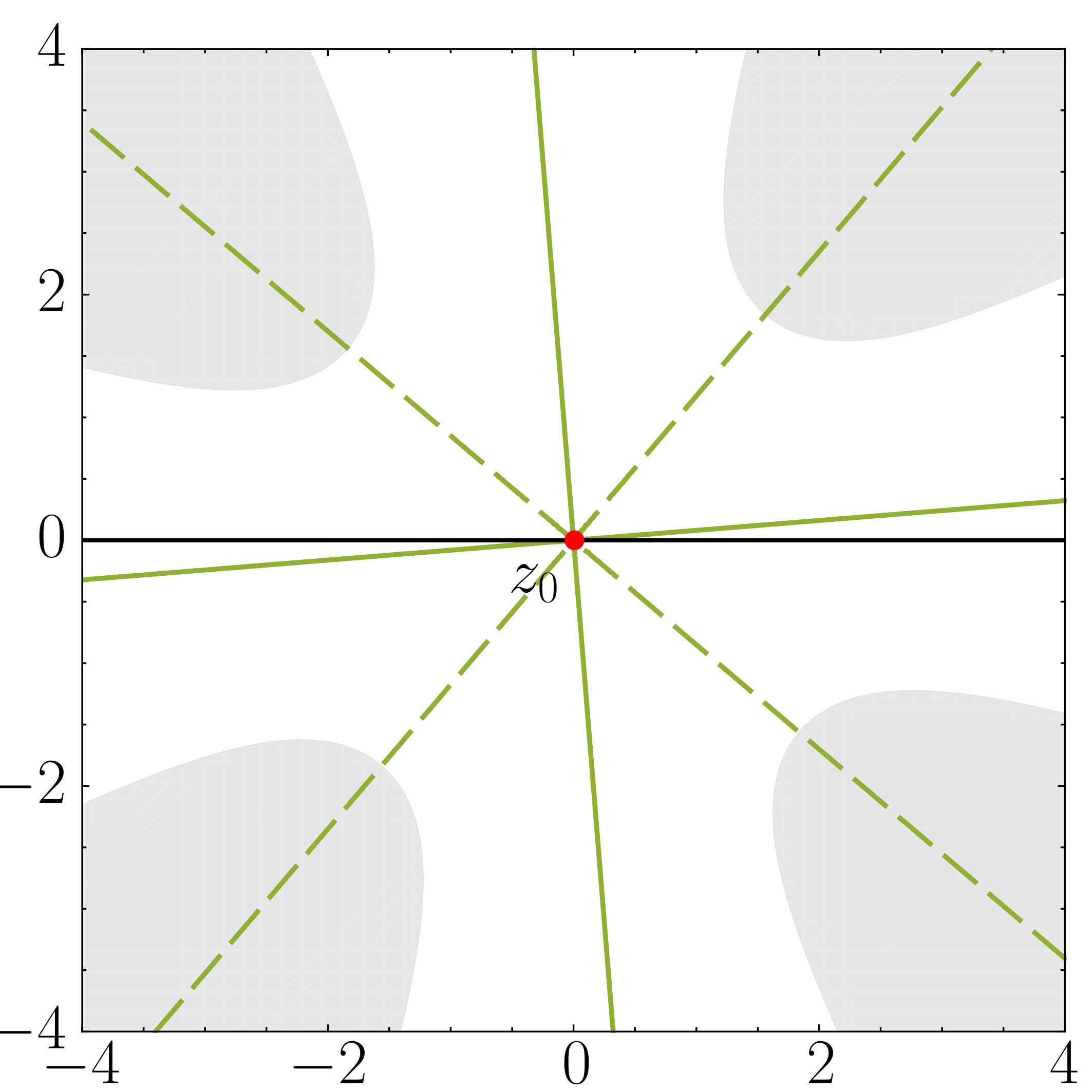}\\
\end{center}
\end{minipage}
%
\begin{minipage}{0.4\linewidth}
\begin{center}
           \includegraphics[width=57mm]{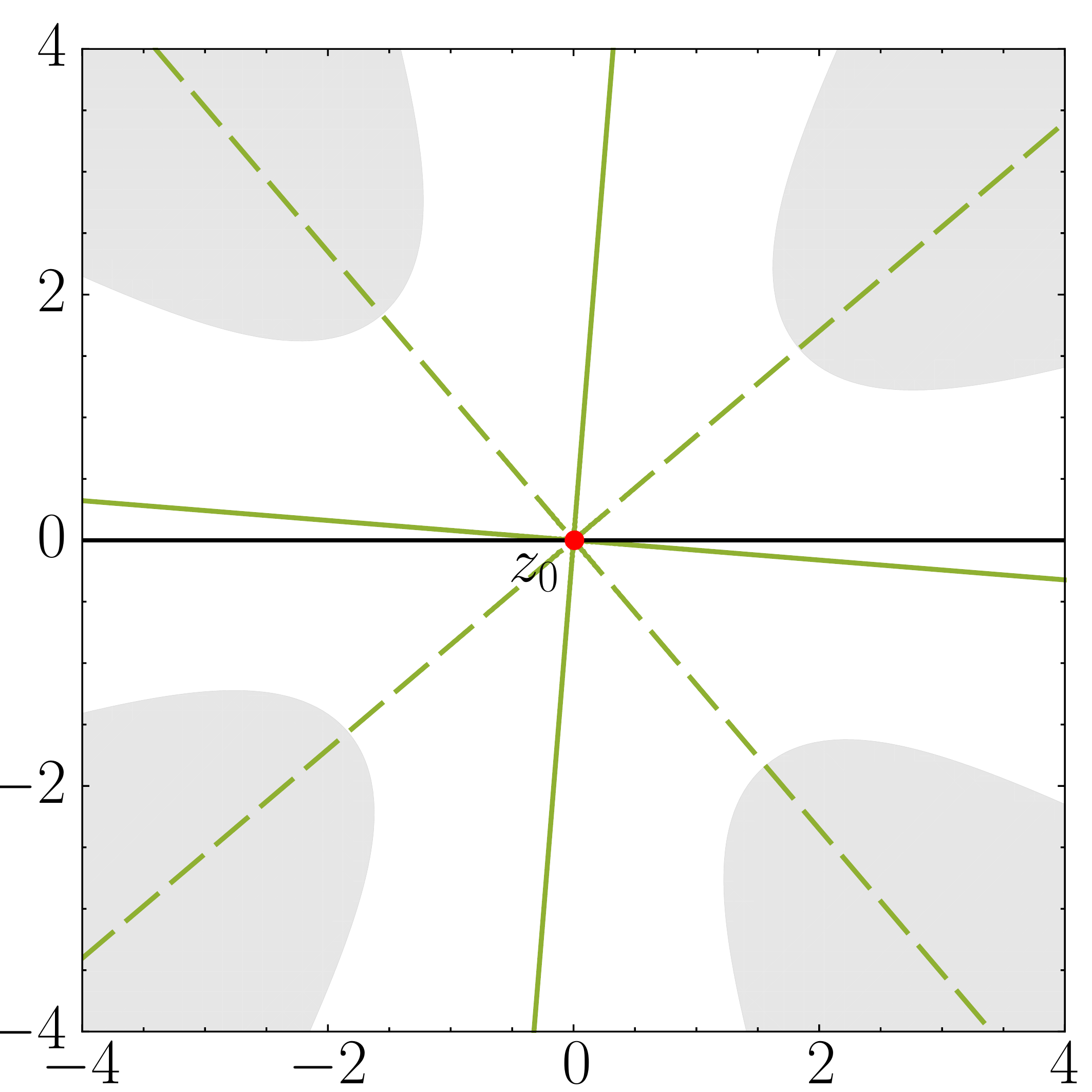}\\
\end{center}
\end{minipage}
\end{center}
\caption{\label{fig:thimbles_x4m1} \small
Downward and upward flows associated to the functions $f(z,1)$ (upper panels), $f(z,-1)$ (middle panels) and $f(z,0)$ (lower panels) in the $z$ plane. The grey sectors correspond to the asymtptotic regions where the integral diverges. The red points are the saddles of the functions $f(z,m)$. Continuous and dashed lines denote downward and upward flows, respectively.  
The lower panels correspond to the degenerate case, where multiple downward and upward flows depart from a saddle point.
We have taken ${\rm Re}\, \lambda=1$, ${\rm Im}\,\lambda >0$ (left panels) and  ${\rm Im}\,\lambda <0$ (right panels).}
\end{figure}

Consider first the case with $m>0$, which, as we will see, is not on a Stokes line for $\lambda$ real and positive. The function $f(z,m)$ has three saddle points: $z_0=0$, $z_{\pm}=\pm i \sqrt{m}$. For real $\lambda$ the upward flows from the saddle $z_0$ hit
the two saddles $z_\pm$. This can be avoided by giving a small imaginary part to $\lambda$
as is shown in fig.~\ref{fig:thimbles_x4m1} (first row) for positive (left) and negative (right) values of Im~$\lambda$. The white regions are those where the integral is asymptotically convergent; by definition, the thimbles (continuous curves) 
start and end in these regions. The upward flows (dashed curves) instead start and end
in the grey regions where the integrand diverges. Notice that the intersection numbers of the upward flows ${\cal K}_\sigma$ with the integration contour are the same in the two cases Im$\lambda\lessgtr 0$ ($n_0=1$, $n_\pm=0$). Therefore, the decomposition (\ref{Cdecomp}) is not ambiguous, 
${\cal C}_x$ coincides with a single thimble and we are not on a Stokes line.

When $m<0$ the integral is on a Stokes line for real positive $\lambda$, since the saddle points are all on the real axis ($z_0=0$, $z_\pm=\pm \sqrt{- m}$). As before the upward flows from $z_0$ hit the other two saddles $z_\pm$, but now the intersection numbers jump across Im~$\lambda=0$ ($n_0=\pm1$, $n_\pm=1$), 
as  can be seen in fig.~\ref{fig:thimbles_x4m1} (second row). Depending on the sign
of Im~$\lambda$ the decomposition (\ref{Cdecomp}) reads
\be \begin{split}
{\cal C}_+ & =  {\cal J}_- -{\cal J}_0 + {\cal J}_+ \,, \qquad {\rm Im}~\lambda > 0 \,, \\
{\cal C}_- & =  {\cal J}_- +{\cal J}_0 + {\cal J}_+ \,, \qquad {\rm Im}~\lambda < 0  \,.
\end{split}
\label{Cx}
\ee
 The integrals over the two paths ${\cal C}_\pm$ coincide when Im~$\lambda\to0$,
as manifest from the figure.

For $m=0$ the only saddle point at $z_0=0$ is degenerate (i.e. $f''(0)=0$) and
multiple upward and downward flows depart from $z_0$ as illustrated in fig.~\ref{fig:thimbles_x4m1} (third row). 
The decomposition rules (\ref{Cdecomp}) do not apply 
and analytic continuation of the parameter $\lambda$ does not help. One possible way
to use saddle point techniques is to define the case $m=0$ 
as the limit  $m\to 0$ of the previous cases, where the three saddle points $z_{0,\pm}$ collide.
An alternative way will be described in sec.~\ref{subsec:EPS}.

\subsection{Asymptotic Series and Borel Sums}

\label{subsec:ASBS}

The integrals $Z_\sigma(\lambda)$ in eq.~(\ref{IgDefTh2}) can be evaluated using saddle-point expansions, that give generically rise to divergent asymptotic expansions.
This is due to the fact that $\lambda=0$ is a singular point, since for any $\lambda<0$ the integral is divergent. 

A series expansion associated to a function $Z(\lambda)$ is asymptotic if, for any fixed order $N$, 
\be
Z(\lambda) - \sum_{n=0}^N Z_n \lambda^n = {\cal O}(\lambda^{N+1}) \,, \quad {\rm as} \quad \lambda\rightarrow 0\,.
\label{ASBS1}
\ee
Different functions can have the same asymptotic expansion,
for instance when the difference is suppressed by a factor $e^{-\alpha/\lambda}$, and hence 
the coefficients of the asymptotic series alone 
do not uniquely fix the function $Z(\lambda)$. 

Divergent asymptotic series provide at best an approximate description 
of the function $Z(\lambda)$, with an accuracy that depends on the behaviour 
of the series coefficients $Z_n$ for $n\gg 1$. Suppose that for $n\gg 1$
\be
Z_n \sim n! a^n n^c \,,
\label{ASBS2}
\ee
for some real parameters $a$ and $c$.\footnote{The analysis that follows can easily be generalized for large-order behaviours of the kind $Z_n \sim (n!)^k a^n n^c$.
In all the cases considered in this paper the parameter $k$ is equal to one.}
The best accuracy for $Z(\lambda)$ is obtained by finding the value $N=N_{{\rm Best}}$ that minimizes the error  $\Delta_Z \sim Z_N \lambda^N$.
Using Stirling formula, one has
\be
N_{{\rm Best}} \approx \frac{1}{|a| \lambda}\,,
\label{ASBS3}
\ee
where the error is asymptotically given by 
\be
\Delta_Z \sim e^{-\frac{1}{|a|\lambda }}\,,
\label{ASBS4}
\ee
independently of $c$ at leading order. This error is consistent with the intrinsic ambiguity related to asymptotic series discussed above.
Keeping more than $N_{{\rm Best}}$ terms in the asymptotic series would lead to an increase in the error.

A possible way to reconstruct a function $Z(\lambda)$  with asymptotic expansion of the form
\be
\sum_{n=0}^\infty Z_n \lambda^n
\label{ASBS5}
\ee
is via Borel resummation. We define the Borel transform
\begin{equation}
{\cal B}Z(t)=\sum_{n=0}^\infty \frac{Z_n}{n!}t^n\,,
\label{BorelSeries}
\end{equation}
which is the analytic continuation of a series with non-zero radius of convergence.\footnote{We assumed here that the coefficients $Z_n$ have the large order behaviour 
given by eq.~(\ref{ASBS2}).}
In the absence of singularities for $t>0$ the integral
\begin{equation} \label{eq:ZB}
Z_{B}(\lambda)=\int_0^\infty dt\, e^{-t} {\cal B}Z(t\lambda)
\end{equation}
defines a function of $\lambda$ with the same asymptotic expansion as $Z(\lambda)$ and
the series (\ref{ASBS5}) is said to be Borel resummable. Since, as we mentioned,
different functions can admit the same asymptotic series,  certain properties of $Z(\lambda)$ and its behaviour 
near the origin have to be  assumed to prove that $Z_B(\lambda)=Z(\lambda)$.\footnote{These assumptions have been given by Watson, see e.g. theorem 136, p.192 of the classic book \cite{Hardy},
and subsequently improved by Nevanlinna, see ref.\cite{Sokal} for a modern presentation.}
These requirements are generically hard to verify. On the other hand, in the specific cases where $Z(\lambda)$ is defined as an integral, one might be able to 
rewrite it directly in the form (\ref{eq:ZB}), so that the equality $Z_B(\lambda)=Z(\lambda)$ can be proved without the need of verifying the above assumptions. 
This is the approach taken in this paper, as we will show in subsec. \ref{subsec:ITBR} and in sec. 3. When $Z_B(\lambda)=Z(\lambda)$ we say that the series 
(\ref{ASBS5}) is Borel resummable to the exact result.

In the following we will be using a generalization of the Borel transform, due to Le Roy, obtained by defining
\be
{\cal B}_bZ(\lambda) \equiv \sum_{n=0}^\infty \frac{Z_n}{\Gamma(n+1+b)} \lambda^n\,,
\label{LeRoy}
\ee
where $b$ is an arbitrary real parameter. The function $Z_B(\lambda)$ now reads
\be
Z_B(\lambda)=\int_0^\infty \! dt \,t^b e^{-t}  \,{\cal B}_bZ(\lambda t)\,,
\label{BorelLeRoy}
\ee
and clearly ${\cal B}_0Z(t) ={\cal B}Z(t)$. Borel-Le Roy trasforms with different $b$ can be related
analytically as follows:
\be \begin{split} \label{eq:Bbrelations}
{\cal B}_b(t)&=t^{-b} \partial_t^{n}\left[t^{b+n} {\cal B}_{b+n}(t)\right] \,, \qquad n\in \mathbb{N}^+ \\
{\cal B}_{b+\alpha}(t)&=\frac{t^{-b-\alpha}}{\Gamma(\alpha)}\int_0^t dt'\,\frac{(t')^b\,{\cal B}_{b}(t')}{(t-t')^{1-\alpha}}\,, \qquad 0<\alpha<1\,.
\end{split} \ee
Note that the position of the singularities of two Borel-Le Roy transforms is the same, which implies 
that Borel summability does not depend on $b$, though the nature of the singularities
might. 

The analytic structure of the Borel transform is connected to the large order behaviour of
the asymptotic series. For example the coefficient $a$ in eq.~(\ref{ASBS2}) determines
the position of the singularity closest to the origin ($\lambda t_\star=1/a$). If $a<0$ the series alternates in sign, the singularity is
on the negative real axis of $t$ and the series is Borel resummable in the absence of further singularities on the positive real axis.
For $a>0$ the closest singularity is on the real axis
and the series is not Borel resummable.\footnote{As we will see, the large-order behaviour of the coefficients $Z_n$ might more generally give rise to poles or branch-cut singularities of ${\cal B}_bZ(t)$ at complex values of $t$. The  conclusion is the same of the case $a<0$.\label{foot1}} 
In this case a lateral Borel resummation can be defined by 
slightly deforming the integration contour of eq.~(\ref{eq:ZB}) 
above or below the singularity.
The resulting ambiguity in the choice of the path is of order 
$e^{-t_\star}=e^{-1/(a\lambda)}$, i.e. ${\cal O}(\Delta_Z)$. 
This ambiguity signals the presence of extra non-perturbative 
contributions to $Z(\lambda)$ not captured by $Z_B(\lambda)$. 
A systematic way of reconstructing the non-perturbative effects from the perturbative
series is the subject of resurgence \cite{ecalle}. 
As we will discuss in detail in the following,
for functions $Z(\lambda)$ of the form (\ref{IgGen})
the deformation defining the lateral Borel resummation corresponds 
to the one needed to avoid Stokes lines in the geometric description 
of subsec. \ref{subsec:LTD}. This leads to non-vanishing intersection numbers (\ref{nsigma}).
For path integrals these numbers are generically infinite but,
luckily enough, we will not need to compute them  
(neither algebraically through resurgence nor geometrically through Picard-Lefschetz theory),
as we will show in the next two subsections.
This is in fact one of the central results of this paper.

\subsection{Borel Summability of Thimbles}

\label{subsec:ITBR}

We saw in subsec.~\ref{subsec:LTD} that the integral $Z(\lambda)$ can be decomposed into a sum
of integrals over thimbles $Z_\sigma(\lambda)$. We will show now that each of these integrals admits
an asymptotic expansion which is Borel resummable to the exact result.

Consider the following change of variable \cite{Berry2,howls}:
\be
t= \frac{f(z) - f(z_\sigma)}{\lambda}\,.
\label{uzChange}
\ee
Recalling eq.~(\ref{udfe2}), we see that for any value of $z$ along ${\cal J}_\sigma$ the variable $t$ is real and non-negative. 
For each value of $t\neq 0$, there are two values $z_{1,2}(\lambda t)\in {\cal J}_\sigma$ satisfying eq.~(\ref{uzChange}):
one for each of the two branches of the downward flow. We take $z_1$ and $z_2$ to be along and opposite to the direction of the thimble.
After this change of variable we get
\begin{equation} \label{eq:simpleBorel}
Z_\sigma(\lambda)=e^{-f(z_\sigma)/\lambda}\, \int_0^\infty dt \, t^{-1/2}\, e^{-t} B_\sigma(\lambda t) \,, \quad
B_\sigma(\lambda t)\equiv \sqrt{\lambda t} \bigg( \frac{g(z_1(\lambda t))}{f'(z_1(\lambda t))}- \frac{g(z_2(\lambda t))}{f'(z_2(\lambda t))}\bigg) \,.
\end{equation}
For small $t$'s, we can expand $f(z)-f(z_\sigma)\propto z^2$ (recall that $f^{\prime\prime}(z_\sigma)\neq 0$) giving $f^\prime(z_{1,2}(t))\propto \sqrt{t} $ so that $B_\sigma(\lambda t)$ is analytic in the origin.\footnote{Note that even if $f''(z_\sigma)=0$ the function $B_\sigma(\lambda t)$ can still be defined in such a way to stay analytic in the origin by rescaling it for a different power of $t$. In particular, if
$f(z(t))-f(z_\sigma)\propto z^n$, with $n>2$, we have  $f^\prime(z_{1,2}(t))\propto t^{1-1/n}$.}
The reader may recognize eq.~(\ref{eq:simpleBorel}) as the Laplace trasform of the
Borel-Le Roy resummation formula (\ref{BorelLeRoy}) with
\be
B_\sigma(\lambda t) = {\cal B}_{-1/2}Z_\sigma(\lambda t)\,.
\label{eq:BcalB}
\ee 
In particular the coefficients of the expansion of $B_\sigma(\lambda)$ around the origin 
are related to those of $Z_\sigma(\lambda)$ by $B_\sigma^{(n)}=Z_\sigma^{(n)}\Gamma(n+1/2)$.
The function $B_\sigma(\lambda t)$ is analytic on the whole semipositive real $t$ axis given
the regularity of $f(z)$ and $g(z)$ and the absence of other saddle points for $f(z)$ 
along the thimble. This proves that the power series of $Z_\sigma(\lambda)$ is Borel resummable.
Not only, but having been able to rewrite the integral directly in terms of a Borel transform of the associated 
asymptotic expansion, we are guaranteed that the Borel resummation reproduces 
the full function $Z_\sigma(\lambda)$.

The original integral (\ref{IgGen}) can then be computed using eq.~(\ref{IgDefTh}) and Borel resummation of the perturbative
expansion of the $Z_\sigma$'s given in eq.~(\ref{eq:simpleBorel}). The contribution associated to the trivial saddle 
(i.e. the one with the smallest $f(z_\sigma)$) can be seen as the perturbative contribution to $Z(\lambda)$, while the other saddles can be interpreted as non-perturbative effects.
When only one saddle contributes, the perturbative contribution is Borel resummable to the exact
result. When more saddles contribute, the perturbative expansion, although Borel resummable, does
not reproduce the full result. If $Z(\lambda)$ is on a Stokes line some of the perturbative expansions 
of the thimbles are not Borel resummable. This is due to singularities of the Borel function induced by the presence of other saddles in the steepest descent path ($f'(z_{1,2}(\lambda t))=0$ for $z\neq z_\sigma$). 

We illustrate the results above using the explicit examples of eq.~(\ref{Ex1}).
We start with the case $m>0$ and, without loss of generality, set $m=1$. 
The original integration path  coincides with the thimble ${\cal J}_0$, 
the only one that contributes, 
and the perturbative expansion is expected to be Borel resummable to the exact result. 
The coefficients $Z^{(m=1)}_{\sigma=0,n}$ of the perturbative expansion of $Z(\lambda,1)$ read
\be
Z_{0,n}^{(1)} = \sqrt{2} (-)^n \frac{\Gamma(2n+\frac 12)}{n!}\,.
\label{series+}
\ee
For large $n$ we have 
\be
Z_{0,n}^{(1)} = (-4)^n \frac{\Gamma(n)}{\sqrt{\pi}}\left (1+{\cal O}\Big(\frac1{n}\Big) \right ).
\label{series+0}
\ee 
The Borel-Le Roy transform (\ref{LeRoy}) with  $b=-1/2$ gives
\be \label{eq:Bm12}
{\cal B}_{-1/2}Z_0^{(1)}(\lambda t) = \sqrt{\frac{1+\sqrt{1+4 \lambda t}}{1+4\lambda t}}\,,
\ee
which presents a branch-cut singularity in the $t$-plane at $\lambda t_\star = -1/4$ but it is regular on the positive real axis. 
By integrating over $t$ one reproduces the exact result (\ref{IgEx}):
\be
 \int_0^\infty \! dt\,   t^{-\frac 12} e^{-t}  {\cal B}_{-1/2}Z_0^{(1)}(\lambda t)  = 
 \frac 1{\sqrt{2\lambda}} e^{\frac{1}{8\lambda}} K_{\frac 14}\Big(\frac{1}{8\lambda}\Big) 
 = Z(\lambda,1)  \,.
\label{BorelIntExp1}
\ee
In this simple case we can also explicitly solve for the change of variable (\ref{uzChange}):
\be
z_{1,2}(\lambda t) = \pm \sqrt{\sqrt{1+4 \lambda t}-1}\,,
\label{Ex3}
\ee
and check  the validity of eq.~(\ref{eq:BcalB}).
The form of ${\cal B}_bZ^{(1)}_0$ depends on the value of $b$.
For instance, the standard Borel function ${\cal B}_0Z^{(1)}_0$ associated to eq.~(\ref{series+}) equals
\be
{\cal B}_0Z^{(1)}_0(\lambda t)
= \sqrt{\frac{8}{\pi}}\frac{K\Big(\frac{-1+\sqrt{1+ 4\lambda t}}{2\sqrt{1+ 4\lambda t}}\Big)}{(1+4\lambda t)^{1/4}}\,,
\label{eq:elliptic}
\ee
where $K(x)$ is the complete elliptic integral of the first kind.
One can check that eq.~(\ref{eq:elliptic}) is in agreement with eq.~(\ref{eq:Bm12}) using the formula~(\ref{eq:Bbrelations}).
We also see, as mentioned, that the position of the singularity of ${\cal B}_0Z^{(1)}_0$ and ${\cal B}_{-1/2}Z^{(1)}_0$ is the same.

The integral with $m<0$ is more interesting because $Z(\lambda,m)$ has a non-trivial
thimble decomposition and is on a Stokes line. 
As we discussed, this is avoided  by taking complex values of $\lambda$. 
Depending on the sign of Im~$\lambda$ the two distinct decompositions in eq.~(\ref{Cx}) are generated. 
Setting $m=-1$ for simplicity and factoring out $e^{-f(z_\sigma)/\lambda}$ from each $Z^{(m=-1)}_\sigma(\lambda)$, the 
coefficients of the perturbative expansions read 
\be
 Z^{(-1)}_{\pm,n} = \frac{\Gamma\left(2n+\frac 12\right)}{n!} \,, \qquad
 Z^{(-1)}_{0,n} = i Z^{(1)}_{0,n} \, \,.
 \label{series-}
 \ee
The Borel-Le Roy transform (\ref{LeRoy}) with  $b=-1/2$ gives
\be
{\cal B}_{-1/2}Z_{\pm}^{(-1)} (\lambda t) = \sqrt{\frac{1+\sqrt{1- 4 \lambda t}}{2(1-4\lambda t)}}\,, \;\; \quad {\cal B}_{-1/2}Z_{0}^{(-1)} (\lambda t) = i {\cal B}_{-1/2}Z_{0}^{(1)} (\lambda t) \,.
\label{B-12}
\ee
The Borel-Le Roy functions ${\cal B}_{-1/2}Z_{\pm}^{(-1)}$ have a branch-cut singularity in the $t$-plane at  $t = 1/(4\lambda)$ and for real positive $\lambda$ the asymptotic series
with coefficients $Z^{(-1)}_{\pm,n}$ are not Borel resummable.
However, the small imaginary part in $\lambda$ needed to avoid the Stokes lines would also allow us to avoid the singularity that now moves slightly below or above the real $t$ axis for Im~$\lambda$ respectively positive or negative. We are effectively performing a lateral Borel summation. After integrating 
over $t$ we get 
\be\begin{split}
 Z^{(-1)}_{\pm}(\lambda) &= {\rm sign(Im}\lambda)\, \frac{i e^{\frac{1}{8\lambda}}}{2\sqrt{\lambda}} K_{\frac 14}\Big(-\frac{1}{8\lambda}\Big)\,, \\
 Z^{(-1)}_{0}(\lambda) &= i Z_0^{(1)}(\lambda)\,.
\end{split} \ee 
Using eq.(\ref{Cx}), the sum of the three contributions  in the limit Im~$\lambda\to 0$ gives
\begin{equation}
 \sqrt{\frac{\pi^2}{4\lambda}} e^{\frac{1}{8\lambda}} \left [ I_{-\frac 14}\left(\frac{1}{8\lambda}\right)+I_{\frac 14}\left(\frac{1}{8\lambda}\right)\right ] =Z(\lambda,-1) \,.
\end{equation}
Notice that the discontinuity of the intersection number $n_0=-{\rm sign(Im}\lambda)$ as Im$\lambda \rightarrow 0$ fixes the ambiguity in the lateral Borel resummation of the
perturbative series around the saddles $z_\pm$.

\subsection{Exact Perturbation Theory}
\label{subsec:EPS}

We have seen in the previous subsections how integrals of the form (\ref{IgGen}) can exactly be computed combining properly resummed
saddle-point contributions. In particular, for real functions $f$, eq.~(\ref{Cdecomp}) is trivial or not
depending on the number of real saddles of $f$.  We will explain in this subsection that the decomposition (\ref{Cdecomp}) in terms of thimbles 
can be modified. This implies that even when $f$ has more real saddles we can trivialize eq.~(\ref{Cdecomp}) so that $Z(\lambda)$ is reproduced by a saddle-point expansion
around one (perturbative) saddle only. This observation will play a crucial role when considering QM, since the computation of the intersection numbers (\ref{nsigma}) is far from trivial
in the path integral case. 

The Lefschetz thimble decomposition associated to the integral (\ref{IgGen}) is governed by the saddle points of $f$ and in particular it is independent of the prefactor $g(x)$.
Define the function
\be
\hat Z(\lambda,\lambda_0) \equiv \frac{1}{\sqrt{\lambda}} \int_{-\infty}^{\infty} \!\! dx\, e^{-\hat f(x)/\lambda} \hat g(x,\lambda_0)\,,
\label{IgDefThM}
\ee
where 
\be
\hat f(x) \equiv f(x) + \delta f(x)\,, \quad \quad \hat g(x,\lambda_0) \equiv g(x) e^{ \delta f (x)/\lambda_0}\,,
\label{EPS1}
\ee
are regular functions of $x$ that satisfy the same conditions as $f(x)$ and $g(x)$, 
in particular\footnote{\label{fn:deltaf} 
It is possible that this condition might be relaxed to some extent. 
It would be interesting to further analyze this point and try to find necessary 
and sufficient conditions for $\delta f(x)$.} 
$\lim_{|x|\to\infty} \delta f(x)/f(x)=0$.
The original integral is recovered by setting $\lambda_0=\lambda$:
\be
\hat Z(\lambda,\lambda)=Z(\lambda)\,. 
\ee
From the point of view of the saddle-point expansion in $\lambda$ at fixed $\lambda_0$, the function $\delta f$ inside $\hat f$ is a ``classical" modification of $f$,
while the factor of $\delta f$ in $\hat g$ is a ``quantum" deformation.
At fixed $\lambda_0$, the thimble decomposition of the integral (\ref{IgDefThM}) is determined by the downward and upward flows associated to the saddle points $z_{\hat\sigma}$ of $\hat f$ and not to  the original saddles $z_\sigma$ of $f$. 
By properly choosing the function $\delta f$, we can generally construct a function $\hat f$ with only one real saddle $x_0$ (for convenience chosen such that $\hat f(x_0) = 0$) 
that trivializes the thimble decomposition to ${\cal C}={\cal C}_x$.
While $Z(\lambda)$ may lie on a Stokes line, so that its perturbation theory is non-Borel resummable and requires extra non-perturbative contributions, 
the asymptotic expansion of $\hat Z(\lambda,\lambda_0)$ in $\lambda$ at fixed $\lambda_0$
will be Borel-resummable to the exact result $\hat Z(\lambda ,\lambda_0)$. Setting then $\lambda=\lambda_0$
allows us to derive the original function $Z(\lambda_0)$.

We call the series expansion of $\hat Z(\lambda,\lambda_0)$ in $\lambda$  at fixed $\lambda_0$ ``exact perturbation theory" (EPT), while we call the ordinary expansion of $Z(\lambda)$  ``standard perturbation theory" (SPT). Note that in general SPT includes both perturbative and non-perturbative saddles.

We illustrate the method by reconsidering the example~(\ref{IgEx}) with $m=-1$, where the contour decomposition (\ref{Cx}) required three different saddle-point contributions. 
Consider the choice 
\be
\delta f(x) = x^2\,,
\ee
so that 
\be
\hat f(x) = \frac 12 x^2 + \frac 14 x^4  = f(x,1)\,, \quad \quad \hat g(x,\lambda_0) = \exp\Big(\frac{x^2}{\lambda_0}\Big)\,.
\ee
The thimble decomposition is now determined by $f(x,1)$, in which case we know that ${\cal C}_x$ coincides with a thimble. The coefficients of the corresponding perturbative expansion read
\be
 \hat Z_n(\lambda_0) = \sqrt{2} (-)^n \frac{\Gamma(2n+\frac 12)}{n!} {}_{1}F_1\Big(-n,\frac 12 -2n;-\frac{2}{\lambda_0}\Big)\,,
 \label{Ex6}
 \ee
where ${}_1F_1(a,b;z)$ is the Kummer confluent hypergeometric function. At any fixed $\lambda_0$, the Kummer function for $n\gg 1/\lambda_0^2$ asymptotes to $\exp(-1/\lambda_0)$ and for large $n$ we have
\be
 \hat Z_n(\lambda_0) \approx  e^{-\frac{1}{\lambda_0}} (-4)^n \frac{\Gamma(n)}{\sqrt{\pi}}\left(1+{\cal O}\Big(\frac{1}{n} \Big)\right) \,,
 \label{Ex6a}
 \ee
where the size of the ${\cal O}(1/n)$ subleading terms depends on $\lambda_0$.
The Borel resummation of the perturbative series  gives
\begin{equation}
{\cal B}_{-1/2}\hat Z(\lambda t,\lambda_0)  =\sum_{n=0}^{\infty} \frac{\hat Z_n(\lambda_0)}{\Gamma(n+1/2)} (\lambda t)^n  \,.
\end{equation}
Recovering the formula for the Borel transform from this equation is non-trivial. We can 
however use eq.~(\ref{eq:simpleBorel}) to get
\begin{equation}
{\cal B}_{-1/2}\hat Z(\lambda t,\lambda_0)  = {\cal B}_{-1/2}Z^{(1)}_0(\lambda t)\
 e^{\frac{\sqrt{1+4 \lambda t}-1}{\lambda_0}}\,, 
 \label{BorelIntExp0M}
\end{equation}
where ${\cal B}_{-1/2}Z^{(1)}_0$ is the Borel-Le Roy function associated to the $m=1$ case, given in eq.~(\ref{eq:Bm12}). 
As expected, no singularities are present on the positive real $t$ axis. 
By taking $\lambda_0=\lambda$ and performing the integral over $t$ one reproduces the exact result for $Z(\lambda,-1)$ given in eq.~(\ref{IgEx}):
\be
\hat Z(\lambda,\lambda) = \int_0^\infty \! dt\, t^{-\frac 12} e^{-t} {\cal B}_{-1/2}\hat Z(\lambda t,\lambda) =
Z(\lambda,-1)  \,.
\label{BorelIntExp1M}
\ee

The above considerations are easily extended to more general functions $f(x)$. 
In particular, for polynomial functions of degree $2n$, independently of the location of the $2n-1$ saddle points and of the corresponding thimble decomposition associated to $f(z)$, we can always
construct a function $\hat f(z)$, for example $\hat f(z)=z^2+f^{(2n)}(0)z^{2n}/(2n)!$,
which has only one real saddle point and a trivial thimble decomposition. 
Notice that the choice of allowed $\delta f(x)$ is arbitrary and all of them are equally good.

Interestingly enough, the method above provides also an efficient way to 
study degenerate cases with $f''(z_\sigma)=0$, where perturbation theory is ill-defined.
Instead of deforming the function $f(z)$, e.g. by adding a small quadratic term $\epsilon z^2$, and 
of analyzing the integral in the limit $\epsilon\rightarrow 0$, we can more simply consider an appropriate function
$\delta f(z)$ that removes the degeneracy, bypassing the need of taking a limit.
For example, consider the integral (\ref{IgEx}) at $m=0$ with the choice
\begin{equation}
\delta f(x)=\frac{x^2}{2}\,,
\end{equation}
so that 
\begin{equation}
\hat f(x)=\frac12 x^2 +\frac14 x^4 = f(x,1)\,, \qquad \hat g(x,\lambda_0)=e^{\frac{x^2}{2\lambda_0}} \,.
\end{equation}
Since this case corresponds to the previous one with $m=-1$ via the rescaling $\lambda_0\to2\lambda_0$, the Borel resummation of the 
perturbative expansion is simply given by ${\cal B}_{-1/2}\hat Z(\lambda t,2\lambda_0)$, with ${\cal B}_{-1/2}\hat Z$ given in eq.~(\ref{BorelIntExp0M}).
Taking $\lambda_0=\lambda$ and performing the integral over $t$, one reproduces the exact result 
\be
 \int_0^\infty \! dt\, t^{-\frac 12} e^{-t} {\cal B}_{-1/2}\hat Z(\lambda t,2\lambda) =
 \frac{\Gamma\left( 1/4 \right ) }{\sqrt2}\, \lambda^{-1/4}  \,.
\ee

\subsection{The Asymptotic Behaviour from Semiclassics}
\label{subsec:ELOB}

The saddle-points, whether or not they contribute to the integral (\ref{IgGen}), dictate the large-order behaviour of the series expansion of adjacent saddles.
In QM this method has been first used\footnote{A similar method was already used in 1964, see ref.\cite{Arkady}. We thank Arkady Vainshtein 
for drawing our attention to his work.} by Bender and Wu in ref.\cite{Bender:1990pd} and extended to QFT by  
Lipatov \cite{Lipatov:1976ny} (see also refs.\cite{Brezin:1976vw,Brezin:1976wa,Brezin:1977gk} for early extensive studies).
For the specific case of finite-dimensional integrals a more rigorous derivation can be found in refs.\cite{Berry2,howls}, where an exact resurgent formula relating the asymptotic series of different saddles has been derived. It has been shown in ref.\cite{Berry2}  that the leading large order behaviour of the coefficients $Z_{\sigma,n}$ is governed by other saddles close to  $z_\sigma$.
More precisely, consider the integral (\ref{IgDefTh2}) as a function of $\lambda=|\lambda|\exp(i\theta)$. The thimble ${\cal J}_\sigma(\theta)$ moves in the complex $z$-plane as $\theta$ is varied. 
For the special values of $\theta$ where the thimble crosses other saddle points the integral 
is on a Stokes line. These saddles are called  ``adjacent" to $z_\sigma$.
Among the adjacent saddles, we denote by $z_{\sigma_0}$ the leading adjacent saddle as the one with the smallest value of $|f(z_{\sigma_0})-f(z_\sigma)|$.
Modulo an overall phase, the large-order behaviour of $Z_{\sigma,n}$ is given by the lowest-order coefficient $ Z_{\sigma_0,0}$ of the series associated to the leading adjacent saddle $z_{\sigma_0}$
 \cite{Berry2}:
\be
Z_{\sigma,n} = \sum_{{z_{\sigma_0}}} Z_{\sigma_0,0}  \frac{(n-1)!}{(f(z_{\sigma_0})-f(z_\sigma))^n} \left(1+{\cal O} \Big(\frac 1n\Big)\right)\,,
\label{ELOB1}
\ee
where $Z_{\sigma_0,0}= g(z_{\sigma_0})/\sqrt{2\pi |f^{\prime\prime}(z_{\sigma_0})|}$
and the sum is present in case we have more than one saddle with the same minimal value of $|f(z_{\sigma_0})-f(z_\sigma)|$. 
Equation (\ref{ELOB1}) justifies and generalizes our working assumption (\ref{ASBS2}) which was valid only for real values of $f(z_{\sigma_0})-f(z_\sigma)$.
Matching the two equations 
we get
\be
a=\frac{1}{f(z_{\sigma_0})-f(z_\sigma)} \,, \quad \quad c = - 1\,.
\label{ELOB2}
\ee
As we mentioned, the coefficient $a$ dictates the location of the leading singularities (i.e. the ones closest to the origin) of the Borel function ${\cal B}Z(t)$.
For real functions $f$ with more than one saddle on the real axis the expansion around a minimum $z_\sigma$ gives $a$ real and positive, 
in agreement with the non-Borel summability of an asymptotic expansion on a Stokes line.\footnote{The argument is valid also when the real saddle  entering eq.~(\ref{ELOB2}) is not the leading adjacent one, in which case the singularity in the positive real $t$ axis will still appear, though it will not be the closest to the origin.}
It is clear from eq.~(\ref{ELOB1}) that in general the Borel function can have leading singularities for complex values of its argument, as anticipated in footnote \ref{foot1}.

The $n$-dependence of the leading large-order behaviour is governed by the function $f(z)$ and is independent of $g(z)$, the latter entering only in the determination of the
overall normalization of the coefficients. For EPT
this implies that at fixed $\lambda_0$, the $n$-dependence of the leading large order behavior of $\hat Z_n(\lambda_0)$ does not depend on $\lambda_0$.
More precisely we have, using eq.~(\ref{ELOB1}),
\be
\hat Z_n(\lambda_0)\approx \sum_{{z_0}} Z_{z_0,0} \frac{(n-1)!}{(\hat f(z_0)-\hat f(x_0))^n} \left(1+{\cal O}\left(\frac1{n}\right)\right)\,,\quad \quad  Z_{z_0,0}= e^{\frac{\delta f(z_0)}{\lambda_0}} \frac{g(z_0)}{\sqrt{2\pi |\hat f^{\prime\prime}(z_0)|}}\,,
\label{ELOB3}
\ee
where $z_0$ are the leading adjacent saddles associated to the (unique) real saddle $x_0$ and $Z_{z_0,0}$ is the leading order term of the series associated to $z_0$.
Given the above choice of $\hat f(z)$, the factor $\hat f(z_0)-\hat f(x_0)$ in eq.~(\ref{ELOB3}) is always either complex or real negative, so that no singularities appear in the positive real $t$ axis of ${\cal B}\hat Z(t)$.
Equation (\ref{ELOB3}) is valid at fixed $\lambda_0$ for parametrically large values of $n$. More specifically we need $n\gg 1$ and $n\gg 1/\lambda_0^2$ in order to suppress the contributions coming from the higher-order coefficient terms $Z_{z_0,1}$, $Z_{z_0,2},\ldots$ associated to the leading adjacent saddle series $Z_{z_0,n}$.
The large-order behaviour (\ref{Ex6a}) is immediately reproduced using eq.~(\ref{ELOB3}).

\section{Path Integrals in QM}
\label{sec:PI}

In this section, after having generalized the results of sec. 2 to higher dimensional integrals, we extend them to path integrals and introduce EPT in QM.

\subsection{Higher Dimensional Integrals}
\label{sec:PLapt}
The analysis of the one-dimensional integral (\ref{IgGen}) performed in sec. \ref{sec:ODI} can be extended to $n$-dimensions. 
Interpreting the domain of integration as an $n$-dimensional cycle ${\cal C}_{\bm x}={\mathbb R}^n$ 
in $n$ complex dimensions (with coordinates ${\bm z}$), like in eq.~(\ref{IgDef}), downward and upward 
flows can be defined generalizing eq.~(\ref{udfe}).
For each saddle ${\bm z}_\sigma$, the Lefschetz thimble ${\cal J}_\sigma$ and its dual cycle ${\cal K}_\sigma$ are obtained  by taking the union of all the down- and up- ward flows. As
for the 1-dimensional case possible Stokes lines can be avoided by taking $\lambda$ complex.
After decomposing the cycle ${\cal C}_{\bm x}$ in terms 
of thimbles, like in eq.~(\ref{Cdecomp}), we are left with the evaluation of integrals of the type
\begin{equation}
Z_\sigma(\lambda)=\lambda^{-n/2}\int_{{\cal J}_\sigma} d {\bm z}\,  g({\bm z}) \, e^{-f({\bm z})/\lambda}\,,
\label{PLapt1}
\end{equation}
with $f$ and $g$ regular functions and such that the integral is convergent for any positive $\lambda$.
By construction the function $f$ has only one non-degenerate saddle $\bm z_\sigma$: 
${\bm \nabla} f({\bm z}_\sigma)=0$ with
$\det[\partial_i \partial_j f({\bm z_\sigma})]\neq 0$. Repeating the same steps as for the one-dimensional case and using known results from
Morse theory, we can prove that  the formal series expansion for $Z_\sigma(\lambda)$ around $\lambda=0$ is Borel resummable to the exact result. 
Indeed performing the change of variables 
\be
 t=\frac{f({\bm z})-f({\bm z}_\sigma)}\lambda\,,
 \label{PLapt2}
\ee
we have (see also refs.\cite{howls,Gukov:2016njj,tHooft})
\be\begin{split}
Z_\sigma(\lambda)&=e^{-f(\bm{z}_\sigma)/\lambda}\int_0^\infty dt\, t^{n/2-1}\,e^{-t}\, B_\sigma(\lambda t)\,,  \\
B_\sigma(\lambda t)&\equiv (\lambda t)^{1-n/2}\int_{{\cal J}_\sigma} d\bm{z} \,  
 g({\bm z})\, \delta[f(\bm{z})-f(\bm{z}_\sigma)-\lambda t]\,.
\end{split}
\label{eq:ZBn}
\ee
\begin{figure}
\centering
\includegraphics[scale=.3]{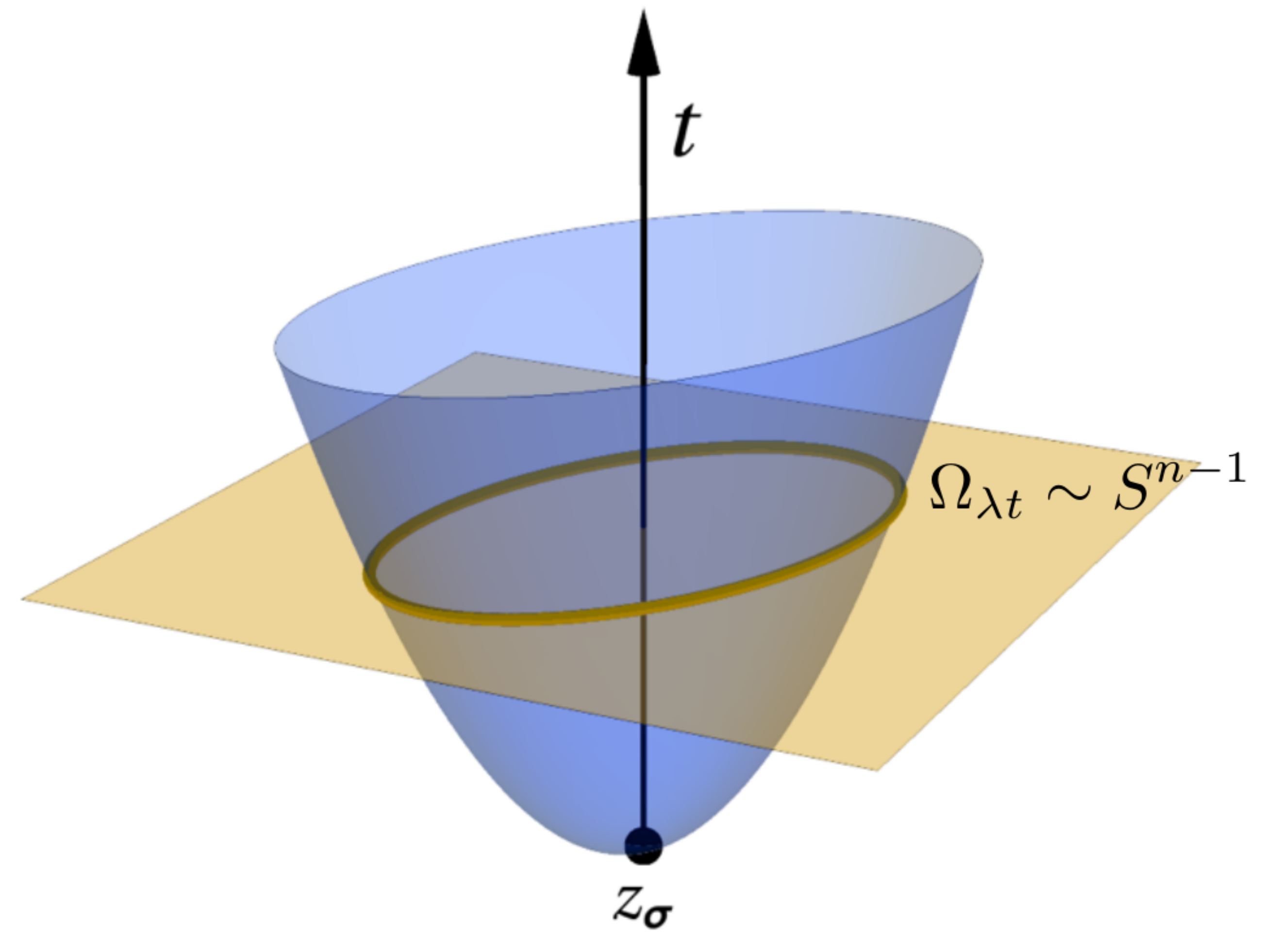}
\caption{\label{fig:thimblendim} \small The support of integration $\Omega_{\lambda t}$ of the Borel
function in $n$ dimensions with topology $S^{n-1}$ is the section of the thimble 
identified by the constraint (\ref{PLapt2}).
}
\end{figure}
The integral in $B_\sigma(\lambda t)$ has support over the  $(n-1)$-dimensional hypersurface
$\Omega_{\lambda t}$, defined by the constraint $f({\bm z})=f({\bm z}_\sigma)+\lambda t$ (see fig.~\ref{fig:thimblendim}).
In the neighborhood of the minimum ${\bm z}_\sigma$ the hypersurface $\Omega_{\lambda t}$ 
has the topology of a compact $n-1$ dimensional sphere $S^{n-1}$.\footnote{In one dimension $S^{n-1}$ reduces to two points, which were denoted by $z_1$ and $z_2$ in eq.~(\ref{eq:simpleBorel}).}
Theorems from Morse theory (see e.g. ref.~\cite{Milnor}) guarantee that this will continue to be true for any 
$t$ as long as no other critical
point of $f({\bm z})$ is met, which is true for thimbles away from Stokes lines. 
Moreover, since $\bm{\nabla}f(\bm{z})\neq \bm{0}$ for $\bm{z}\neq \bm{z}_\sigma$,
 it follows that the integral defining $B_\sigma(\lambda t)$ is finite for any value of $t>0$.
Similarly to the one-dimensional case, for $t\to 0$ one has $\bm{\nabla}f(\bm{z})={\cal O}(t^{1/2})$.
Taking into account the $\lambda t$ dependence from the volume form we see that 
$B_\sigma(\lambda t)$ is analytic in the whole semipositive real axis including the origin.
We conclude that the power series of $Z_\sigma(\lambda)$ in $\lambda$ is Borel
resummable to the exact result and $B_\sigma(t) = {\cal B}_{n/2-1}Z_\sigma(t)$.
Depending on whether $n$ is even or odd, the first relation in eq.~(\ref{eq:Bbrelations}) 
allows us to rewrite Borel and $b=-1/2$ Borel-Le Roy transforms 
as simple derivatives of the formula above, namely 
\begin{equation} \begin{split} \label{eq:Bnrel}
{\cal B}_0Z_\sigma(t)&=\partial_t^{k-1} \int_{{\cal J}_\sigma}\!  d\bm{z} \,  
 g({\bm z})\,  \delta[f(\bm{z})-f(\bm{z}_\sigma)- t]\,, \qquad n=2k\,, \\ 
{\cal B}_{-1/2}Z_\sigma(t)&=  \sqrt t \,  \partial_t^{k} \int_{{\cal J}_\sigma}\!  d\bm{z} \,  
 g({\bm z})\,\delta[f(\bm{z})-f(\bm{z}_\sigma)- t] \,, \qquad n=2k+1 \,.
\end{split} \end{equation}

\subsection{The Lefschetz Thimble Approach to the Path Integral}

We are now ready to discuss path integrals in QM. Consider
\begin{equation}
{\cal Z}(\lambda) =\int {\cal D}x(\tau)\, G[x(\tau)] e^{-S[x(\tau)]/\lambda}\,,
 \label{PLapt3}
\end{equation}
with $\tau$ the Euclidean time. The action $S$ is of the form
\begin{equation}
S[x]=\int d\tau\, \left[\frac12 \dot x^2+V(x) \right]\,,
 \label{PLapt4}
\end{equation}
where the potential $V(x)$ is assumed to be an analytic function of $x$, with
$V(x)\to\infty$ for $|x|\to \infty$, so that the spectrum is discrete.
In analogy to the finite dimensional cases the functionals $S[x]$ and $G[x]$ 
are regular and such that the path integral is well-defined for any $\lambda\geq 0$. 
The measure ${\cal D}x$ includes a $\lambda$-dependent normalization factor 
to make ${\cal Z}(\lambda)$ finite. The integration is taken over real $x(\tau)$ configurations
satisfying the boundary conditions of the observable of interest.

By definition the path integral is the infinite dimensional limit 
of a finite dimensional integral, namely eq.~(\ref{PLapt3}) means
\begin{equation} \label{eq:ZlambdaN}
{\cal Z}(\lambda) = \lim_{N\to \infty} \int {\cal D}^{(N)}x(\tau)\, G^{(N)}[x(\tau)] e^{-S^{(N)}[x(\tau)]/\lambda}\,,
\end{equation}
where the limit is now explicit and $G^{(N)}[x(\tau)]$, 
$ S^{(N)}[x(\tau)]$, $ {\cal D}^{(N)}x(\tau)$, 
are discretized versions 
of the functionals  $G$ and $S$ and of the path integral measure,
which includes the factor $\lambda^{-N/2}$.
Such limit can be twofold: The continuum limit, always present, and the infinite time limit,
relevant for the extraction of certain observables. The former is not expected to generate
problems in QM, since after having properly normalized the path integral all quantities
are finite in this limit. The infinite time limit could 
instead be more subtle and we will discuss later when it may lead to troubles.
For the moment we restrict our analysis to path integrals at finite time so that
the limit in eq.~(\ref{eq:ZlambdaN}) only refers to the continuum one.

Similarly to the finite dimensional case, the first step is to identify all the (real and complex) saddles $z_\sigma(\tau)$ (the solutions of the equations of motion) 
of the action $S[x]$ and to construct the analogue of the upward and downward
flows from each $z_\sigma(\tau)$. Given the infinite dimensional nature of the path integral,
the number of saddles is also infinite.
In general a systematic analysis and computation of all the relevant flows and 
intersection numbers is impractical.  In specific cases, however, only a few real 
saddle point solutions exist and we may hope to reconstruct the full answer from a finite set
of saddle point expansions. In particular, if the equations of motion admit only one real
solution, the domain of the integration (all real paths satisfying the boundary conditions) 
coincides with a thimble. We will now show that such path integral (and similarly any path
integral over thimbles) admits a  perturbation theory  which is Borel resummable to the exact result.

The integral inside the limit in eq.~(\ref{eq:ZlambdaN}) 
is finite dimensional and can now be treated as before, in
particular we can rewrite it using eqs.~(\ref{eq:ZBn}) and (\ref{eq:Bnrel}) as
\begin{equation} \begin{split} \label{eq:ZBn2}
{\cal Z}(\lambda) & = \lim_{N\to \infty} e^{-\frac{S^{(N)}[x_0(\tau)]}\lambda}\int_0^\infty dt\, t^{-1/2}e^{-t} {\cal B}_{-1/2}^{(N)} {\cal Z}(\lambda t) \\
{\cal B}_{-1/2}^{(N)} {\cal Z}(\lambda t) & = \sqrt{\lambda t}\  \partial_{\lambda t}^{N}
 \int {\cal D}^{(N)}_{\lambda=1}x(\tau)\, G^{(N)}[x(\tau)]\,  \delta[S^{(N)}[x(\tau)]-S^{(N)}[x_0(\tau)]-\lambda t] \,,
\end{split} \end{equation}
where for definiteness we discretized the path integral into a $2N+1$ dimensional one and ${\cal D}_{\lambda=1}^{(N)}x(\tau)$ is the discretized measure without the $\lambda$ dependence (i.e. with $\lambda=1$). The regularity of the functionals $S$ and $G$ and the absence of other real saddle points allow a choice of discretization for which
the Borel-Le Roy function ${\cal B}_{-1/2}^{(N)} {\cal Z}(\lambda t)$
is finite and integrable for any $N$. 
In QM the absence of divergences in the continuum limit 
strongly suggests that the exchange of the limit with the integral
in the first line of eq.~(\ref{eq:ZBn2}) can be performed safely.
The function ${\cal B}_{-1/2}^{(\infty)} {\cal Z}(\lambda t)$ will then correspond to the Borel-Le Roy transform in the continuum limit, 
which will be integrable and will reproduce the full result ${\cal Z}(\lambda)$.
As a check we verified the finiteness of ${\cal B}_{-1/2}^{(\infty)} {\cal Z}(\lambda t)$ at $\lambda t=0$ 
(which reproduces the known results for the harmonic oscillator)
as well as at any order in $\lambda t$ in polynomial potentials (see the appendix).

We are then led to the following result:

{\em If the action $S[x(\tau)]$ has only one real saddle point $x_0(\tau)$ satisfying the boundary conditions
implicit in eq.~(\ref{PLapt3}), such that $\det S''[x_0(\tau)]\neq 0$, then no thimble decomposition is needed and
the formal series expansion of ${\cal Z}(\lambda)$ around $\lambda=0$, corresponding to the saddle point expansion of the path integral, is Borel resummable to the exact
result.}

If the action $S[x]$ admits one real saddle only, in general it will admit several (or an infinite number of) complex saddles (or complex instantons).
All these complex instantons, however, do {\it not} contribute to the path integral. Analogously to the finite-dimensional cases, whenever more than one real saddle point with finite action satisfying the boundary conditions of the path integral exists, the perturbative series generically will not be Borel resummable, as a result of the Stokes phenomenon. 

\begin{figure}
\begin{center}
\includegraphics[scale=.4]{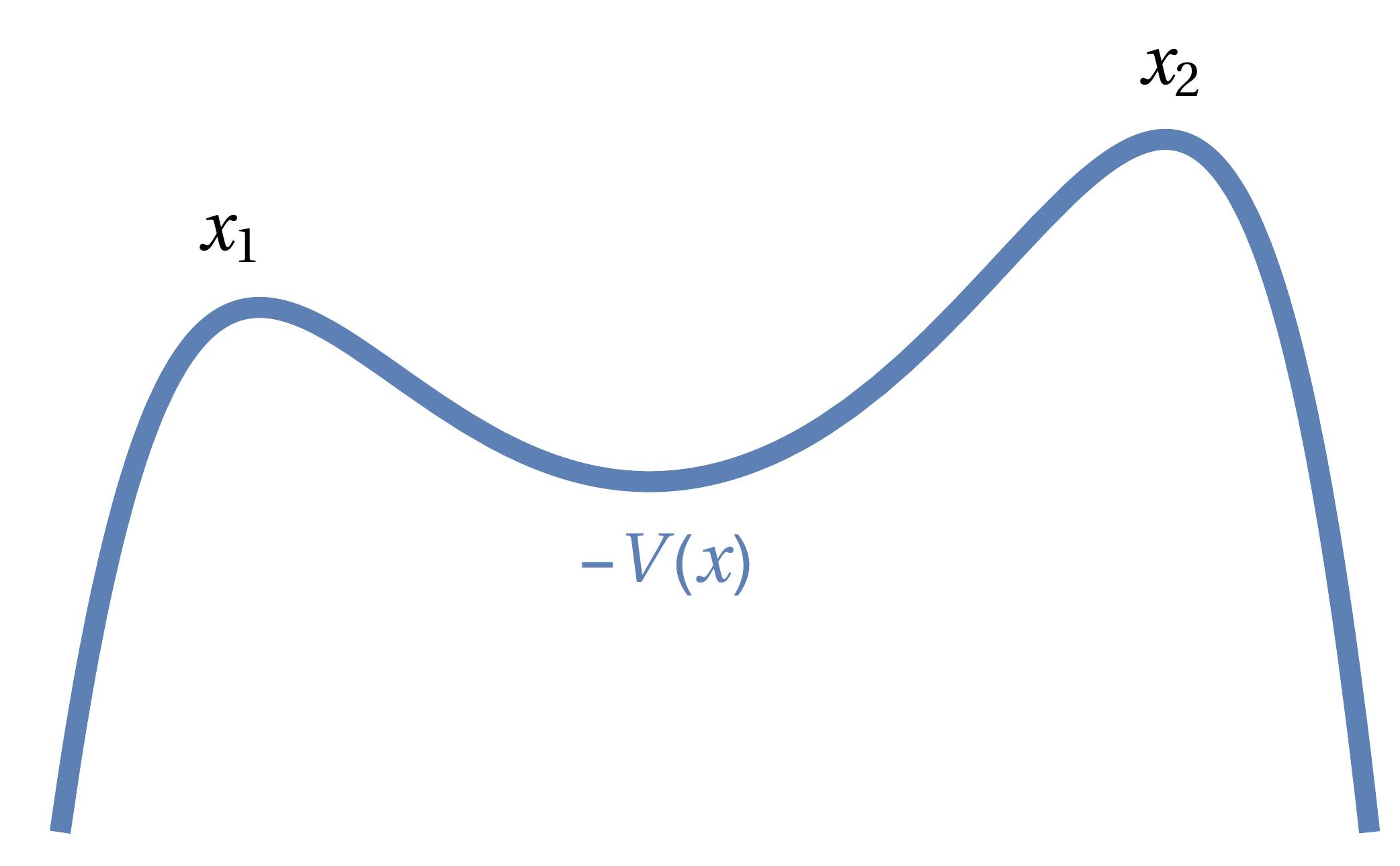}\\
\end{center}
\caption{\small 
Sketch of a bounded (inverted) potential with more than one critical point.}
\label{fig:pot}
\end{figure}

Boundary conditions determine the number of real saddle points of $S$ and hence are of crucial importance
to check the validity of our working assumption.
 As a result the same theory may have some observables that are Borel resummable to the exact result and some for which the perturbative series requires the inclusion of non-perturbative effects. 
It might be useful to illustrate the point with an example.
Consider a QM system with an (inverted) potential like the one depicted in fig. \ref{fig:pot} and define
\begin{equation}
{\cal W}(\lambda,\beta,x_0)=
\int_{x(\beta/2)=x(-\beta/2)=x_0}
\hspace{-70pt} {{\cal D}x(\tau)\, e^{-S[x(\tau)]/\lambda}} \ 
= \sum_k |\psi_k(x_0;\lambda)|^2 e^{-\beta E_k(\lambda)}\,,
 \label{PLapt3d}
\end{equation}
where $E_k(\lambda)$ and $\psi_k(x;\lambda)$ are the eigenvalues and the eigenfunctions of the system, respectively.
Depending on $x_0$, the action $S$ admits one or more real saddle points.
For instance, for $x_0>x_2$ only one real solution exist. For $x_0<x_2$ depending
on $\beta$ one or more real solutions are allowed. 

The partition function is related to ${\cal W}(\lambda,\beta,x_0)$ by 
\be
{\cal Z}(\lambda,\beta) = \int_{-\infty}^\infty \! dx_0 \, {\cal W}(\lambda,\beta,x_0) 
= \sum_n e^{-\beta E_n(\lambda)}\,,
 \label{PLapt3b}
\end{equation}
which corresponds to summing over all real periodic trajectories and it is not Borel resummable. 

We now discuss the infinite $\beta$ limit, which is relevant for the extraction of
some observables such as the properties of the ground state
(eigenvalue, eigenfunction, \dots). Unlike the continuum limit, the large-$\beta$ limit
generically does not commute with the thimble decomposition. There are cases where
the path integral admits more than one real saddle at finite $\beta$ but
only one survives at $\beta\to \infty$. There are other cases instead where only
one real saddle exists for any finite $\beta$ but the path integral lies on a Stokes
line at $\beta=\infty$.

Consider for instance the ground state energy
\be
E_0(\lambda) = -\lim_{\beta \rightarrow \infty} \frac1{\beta}\log {\cal Z}(\lambda,\beta) \,.
\ee
For the example of tilted double-well potential discussed before,
${\cal Z}(\lambda,\beta)$ has multiple real saddles for any finite $\beta$, corresponding to solutions of the equations of motion with period $\beta$. Besides the trivial one, 
$x(\tau)=x_2$, the leading saddle corresponds to the solution 
$x(\tau)=x_1$, which is suppressed by a factor $e^{-\beta [V(x_1)-V(x_2)]/\lambda}$. 
Therefore in the limit $\beta\to\infty$ only
the thimble associated to the true minimum gives a non-vanishing contribution. 
The perturbative series for $E_0$ is then Borel resummable to the full answer,
though ${\cal Z}(\lambda,\beta)$ at finite $\beta$ is not. 
This result is more manifest if we use the alternative formula 
\be
E_0(\lambda) = -\lim_{\beta \rightarrow \infty} \frac1{\beta}\log {\cal W}(\lambda,\beta,x_2)  \,.
\ee
Since ${\cal W}(\lambda,\beta,x_2)$ has only one saddle ($x(\tau)=x_2$) for any $\beta$,
the Borel summability of $E_0(\lambda)$ follows trivially from our analysis.\footnote{
The Borel summability of
${\cal W}(\lambda,\beta,x_2)$ for any $\beta$ and its explicit form (\ref{PLapt3d}) suggest 
that the same conclusion should hold for the rest of the spectrum.}

The same discussion applies for any bound-state potential with a unique global minimum. 
When the minimum is not unique generically the perturbative series of $E_0$ is not Borel resummable 
because in the large $\beta$ limit an infinite  number of saddles with finite action are present, 
independently of the functional ${\cal W}$ used. This is also true if the degeneracy of the absolute
minimum is lifted at the quantum level, i.e. $V(x_1)-V(x_2)={\cal O}(\lambda)$,
as it will become more clear below.
We will discuss in more detail the properties of the Borel transform of $E_0$ for the different cases in sec.~\ref{sec:examples}.

The analysis is particularly simple for potentials $V(x)$ that have 
a single non-degenerate critical point (minimum). Without loss of generality
we take it to be at the origin with $V(0)=0$. Indeed, independently of 
the boundary conditions, there is always a single real saddle point
both at finite and infinite $\beta$. Since our considerations apply 
for any allowed choice of the functional $G[x]$ in eq.~(\ref{PLapt3})
we are led to argue that perturbative series of any observable
is Borel resummable to the exact result. By \emph{any observable} we mean any path integral with regular boundary conditions and analytic functions of them, such as partition functions, energy eigenvalues, eigenfunctions, etc.
In this way we recover in a simple and intuitive way known results 
\cite{Loeffel:1970fe,Simon:1970mc,Graffi:1990pe} 
on the Borel summability of the energy spectrum of 
a class of anharmonic potentials and extend them to more 
general QM systems and observables.

\subsection{Exact Perturbation Theory}

Interestingly enough, the method of subsec. \ref{subsec:EPS} can easily be extended to the QM path integral (\ref{PLapt3}). 
Suppose we can split the potential $V=V_0+\Delta V$ into the sum of 
two potentials $V_0$ and $\Delta V$ such that\footnote{The second condition may be too conservative, see also footnote \ref{fn:deltaf}.}  
\begin{enumerate}
\item $V_0$ has a single non-degenerate critical point (minimum);
\item $\lim_{|x|\rightarrow \infty} \Delta V/V_0= 0$\,.
\end{enumerate}
Consider then the auxiliary potential 
\be
\hat V = V_0+ \frac{\lambda}{\lambda_0}  \Delta V \equiv 
V_0+ \lambda V_1 \,, 
\ee
where $\lambda_0$ is an arbitrary positive constant and
define the modified path integral
\begin{equation}
 \hat{\cal Z}(\lambda,\lambda_0)=\int {\cal D}x\ G[x]\ e^{-\frac{\int \!d\tau \,\Delta V}{\lambda_0} }  
\ e^{-\frac{S_0}{\lambda}}\,, \qquad 
S_0 \equiv   \int \!d\tau \left[\frac12 \dot x^2+V_0 \right].
 \label{PLapt5}
\end{equation}
Since $\hat{\cal Z}(\lambda,\lambda)= {\cal Z}(\lambda)$, the latter
can be obtained by the asymptotic expansion in $\lambda$ of $\hat {\cal Z}$ (EPT), 
which is guaranteed to be Borel resummable to the exact answer.

We can then relax the requirement of a single critical point and state our general result:

{\it All observables in a one-dimensional QM system with a bound-state potential $V$ for which points 1. and 2. above apply  are expected to be entirely reconstructable from a single perturbative series.}

Generally the decomposition $V = V_0+\Delta V$ is far from being unique. EPT is defined as an expansion around the minimum of $V_0$,
which does not need to be a minimum (or a critical point) of the original potential $V$. The number of interaction terms present in EPT also depends on the particular decomposition performed
and in general is higher than the number of interaction terms present in the original theory (SPT). Since the mass term we choose for $V_0$ might differ from the one of $V$, so will do the effective
couplings of SPT and EPT.  As long as conditions 1. and 2. above are fulfilled, any choice of EPT is equivalent to any other, though
in numerical computations with truncated series some choices might be more convenient than others. 

The leading large-order behaviour of the coefficients associated to the asymptotic expansion of the ground state energy 
associated to $\hat {\cal Z}(\lambda,\lambda_0)$ can be deduced using the results of refs.\cite{Lipatov:1976ny,Brezin:1976vw,Brezin:1976wa}
(see in particular sec. II of ref.\cite{Brezin:1976wa}). The large-order behaviour of the ground state energy coefficients is governed by the action $S_0$ only. 
In analogy to the one-dimensional integral, the functional $G\exp(-\int \Delta V/\lambda_0)$ in eq.~(\ref{PLapt5}) governs only the overall $n$-independent size of the coefficients. 

Note that so far we used non-canonical variables with the coupling constant $\lambda$ playing the role of $\hbar$---the saddle-point expansion is the loopwise expansion. This means that in the canonical basis the potential $V(x)$ turns into $V(x;\lambda)$ defined as
\be
V(x;\lambda) = \frac{V(\sqrt{\lambda}\, x)}{\lambda}\,.
\label{Ex0}
\ee
On the other hand, the coupling constant dependence of a generic QM potential 
may not be of the form in eq.~(\ref{Ex0}). For example, the expansion in $g$ for the potential 
\be
V(x;g) = x^2 + g x^4 +  g x^6
\ee
does not correspond to the loopwise parameter. Indeed, setting  $\lambda=\sqrt g$, the terms $x^2+\lambda^2 x^6$ satisfy eq.~(\ref{Ex0}) while the term $\lambda^2 x^4$ is effectively a one-loop potential that should be included in the functional
$G[x]$ of eq.~(\ref{PLapt3}).

\section{Quantum Mechanical Examples}
\label{sec:examples}

In this section we study numerically polynomial QM systems 
in SPT and EPT, providing extensive evidence of the results obtained in the previous sections.
A summary of part of these results appeared in ref.\cite{shortpaper}.

The perturbative series in both SPT and EPT are obtained by using the Mathematica \cite{mathematica} package ``BenderWu" of ref.\cite{Sulejmanpasic:2016fwr} which 
is based on recursion relations among the perturbative coefficients first derived by Bender and Wu in refs.\cite{Bender:1971gu,Bender:1990pd}. We consider up to $N$ orders in the perturbative expansion (for
EPT we fix the auxiliary parameter $\lambda_0$ to a given value) and we approximate the Borel function with Pad\'e approximants.\footnote{We also considered other approximation methods, such as the conformal mapping of refs.\cite{loeffel, LeGuillou:1979ixc}. While the results are consistent with those obtained using Borel-Pad\'e approximants, the latter typically give a better numerical precision for $N\gg 1$. On the other hand, at small $N$
the conformal mapping method is more reliable because of numerical instabilities of the Borel-Pad\'e approximants.}  
For definiteness we use the Borel-Le~Roy function  ${\cal B}_{b}{\cal Z}(\lambda)$ with $b=-1/2$,
which numerically seems a convenient choice leading to more accurate Pad\'e approximants.
The numerical computation of the integral in eq.~(\ref{BorelLeRoy}) gives the final result (evaluated for the value of the coupling $\lambda=\lambda_0$ in EPT).\footnote{Results with $N=100\div 500$ 
are obtained within minutes$\div$hours with a current standard laptop.}
In the following we will refer to the above procedure as the Pad\'e-Borel method.
The result obtained is then compared with other numerical methods such as the 
Rayleigh-Ritz (RR) method (see e.g. ref.\cite{vieira} for 
some explicit realizations). For polynomial potentials of small degree an efficient implementation
is as follows: One starts from the truncated basis $|k_0\rangle$, $k=1,\ldots, N_{RR}$ of the 
harmonic oscillator eigenfunctions, and then computes the full Hamiltonian matrix $H_{kh} 
= \langle k_0| H | h_0\rangle$, which is almost diagonal.
The approximate energy levels and eigenfunctions of the system are given by the eigenvalues and eigenvectors of $H_{kh}$. This method converges to the exact result very quickly. 
The accuracy depends on $N_{RR}$ and on the energy level considered. 
The lower is the level, the higher is the accuracy.

We mostly focus on the energy eigenvalues $E_k(\lambda)$, though the eigenfunctions 
$\phi_k(x;\lambda)$ are also considered. 
Since the package \cite{Sulejmanpasic:2016fwr} computes non-normalized wavefunctions, we define
\be
 \psi_k(x;\lambda) \equiv \frac{\phi_k(x;\lambda)}{\phi_k(x_0;\lambda)}
\label{eq:WFx0}
\ee
for some $x_0$ and compute
\be
\Delta \psi_k(x;\lambda) = \frac{ \psi_k^{\rm{RR}}(x;\lambda)- \psi_k^{\rm{EPT}}(x;\lambda)}{ \psi_k^{\rm{RR}}(x;\lambda)}\,,
\label{eq:WF}
\ee
where for simplicity we omit the $x_0$ dependence in the $\psi_k(x;\lambda)$.

\subsection{Tilted Anharmonic Oscillator}
\label{subsec}

\begin{figure}[t]
\centering
\hspace{-1cm}
\includegraphics[scale=.4]{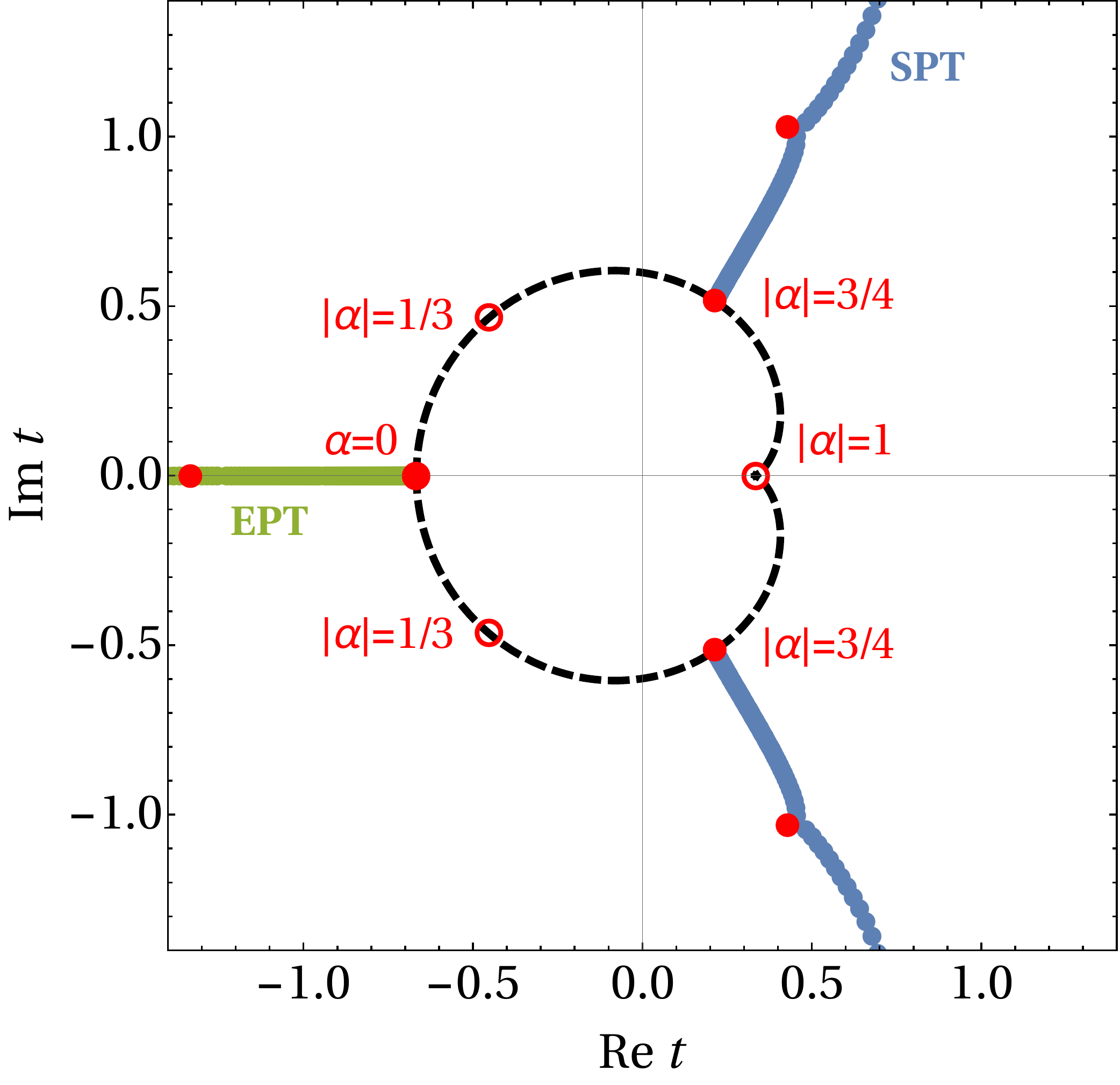}
\caption{\small Location of the singularities in the  Borel plane for the standard and exact perturbative series of $E_0(\lambda,\alpha)$ for the potential (\ref{eq:V-TiltedAnh}). The dashed line 
represents the location of the leading singularities as expected from eq.~(\ref{complexIns}) with $|\alpha|\in[0,1]$. 
The red bullets indicate the position of the first and second complex instantons for $|\alpha|=3/4$. The regions where the simple poles of the Pad\'e-Borel approximants accumulate are depicted in blue and green.}
\label{fig:BorelPlane}
\end{figure}
\begin{figure}[h]
\centering
\includegraphics[scale=.26]{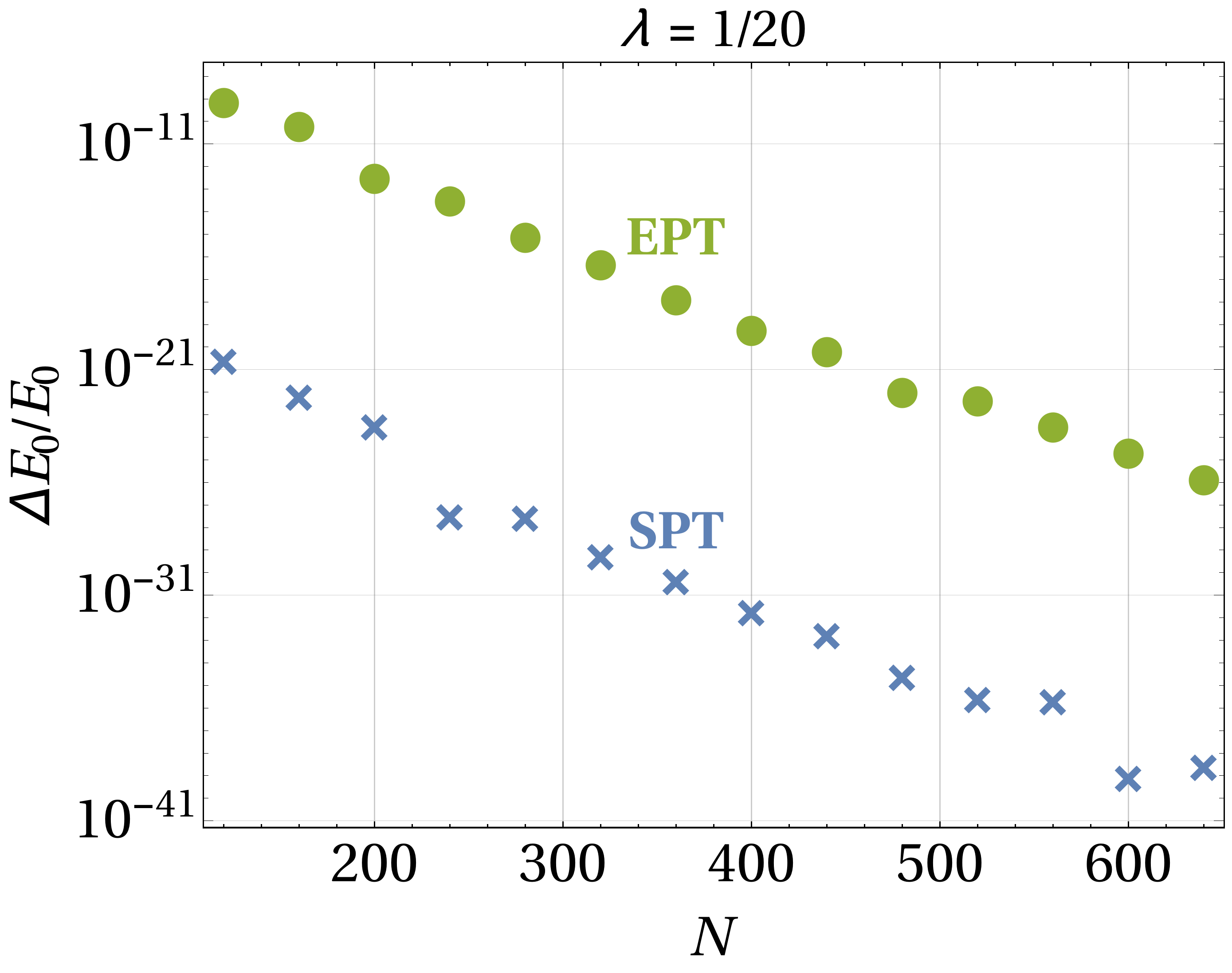}\qquad
\includegraphics[scale=.26]{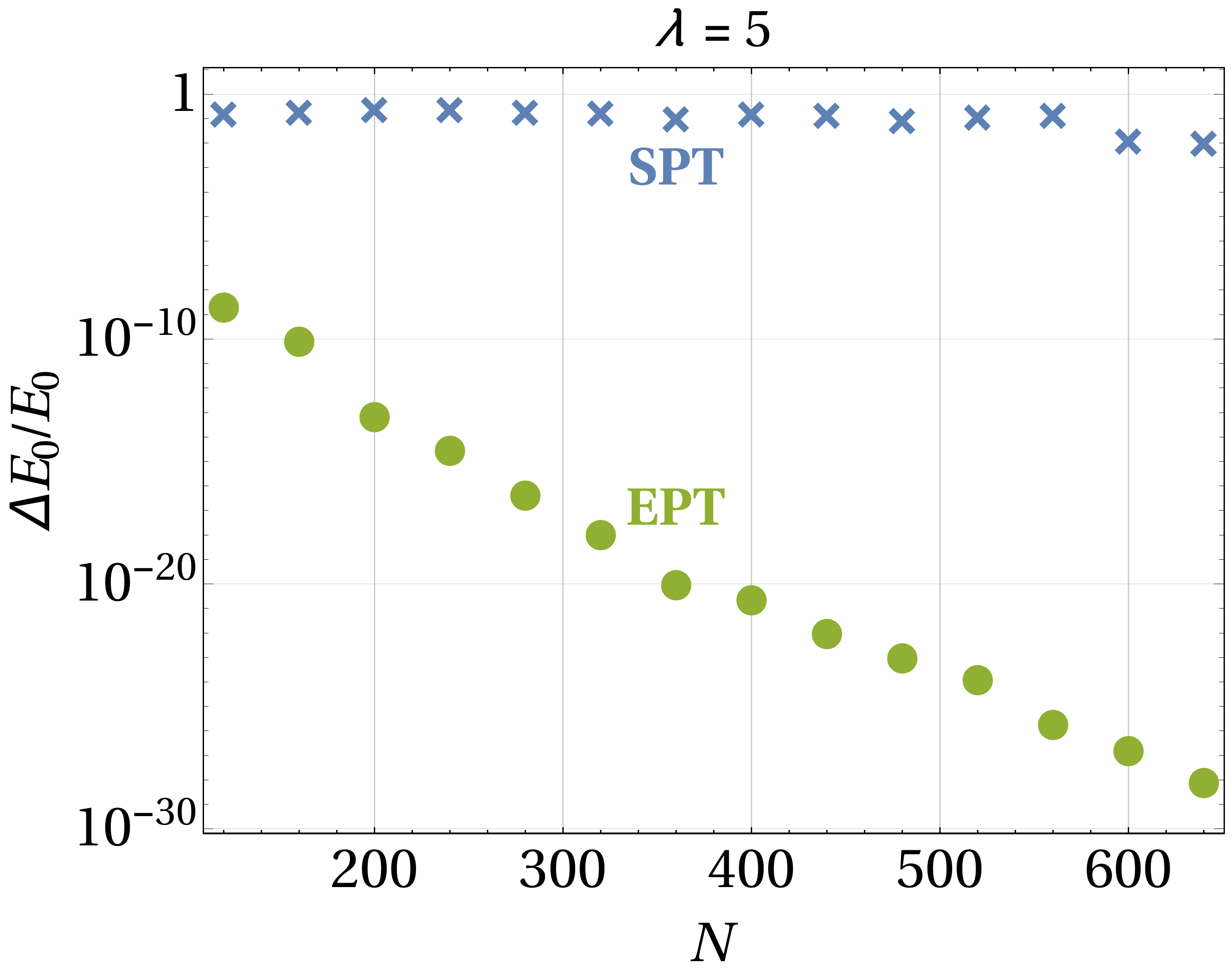}
\caption{\small Comparison between the relative error $\Delta E_0/E_0$ in the computation of the ground state energy using Borel-Pad\'e approximants of the series coming from 
eqs.~(\ref{eq:V-TiltedAnh}) (SPT) and (\ref{eq:V-TiltedHatAnh}) (EPT) as a function of the number $N$ of series coefficients retained. (Left panel) weak coupling $\lambda=1/20$, (right panel) strong coupling $\lambda=5$.}
\label{fig:TA-accuracy}
\end{figure}

The first example we consider is the tilted anharmonic oscillator
\be
V(x;\lambda)=\frac{1}{2}x^2 +\alpha \sqrt{\lambda} x^3+ \frac{\lambda}{2}x^4 \,,
\label{eq:V-TiltedAnh}
\ee
where $\alpha$ is a real parameter.  For $|\alpha|<2\sqrt{2}/3$, the potential has a unique minimum at $x=0$.
According to our results, SPT is then Borel resummable to the exact value for all the observables.
For $2\sqrt{2}/3\leq |\alpha|<1$ some quantities, such as the ground state energy $E_0(\lambda,\alpha)$, are Borel resummable to the exact value
while some are not, such as the partition function. The cases $|\alpha|=1$ (symmetric double well) and $|\alpha|>1$ (false vacuum) will be discussed in the next subsections.

For definiteness, let us look at the ground state energy $E_0(\lambda,\alpha)$.
The position of the leading singularities in the associated Borel plane is dictated by the value of the action $S[z_\pm]$ on the nearest saddle points, which for $|\alpha|<1$ are complex instantons $z_\pm$  \cite{Brezin:1976wa}: 
\be
\lambda t_\pm = \lambda S[z_\pm]= -\frac 23+\alpha^2+\frac 12 \alpha (\alpha^2-1)\Big( \log\frac{1-\alpha}{1+\alpha} \pm i \pi \Big)\,.
\label{complexIns}
\ee
This expectation is confirmed by a numerical analysis with Borel-Pad\'e approximants (see  fig.~\ref{fig:BorelPlane}).
The ground state energy coefficients $E_{0,n}(\alpha)$ for $n\gg 1$ oscillate with a period given by $2\pi/|{\rm Arg} \, t_\pm|$.
As long as the coefficients oscillate, the observable is Borel resummable. The period is minimum at $\alpha=0$, where the coefficients alternate, 
and grows with $|\alpha|$ until it becomes infinite at $|\alpha|=1$ and Borel resummability is lost. In numerical evaluations at fixed order $N$ the best accuracy is obtained at $\alpha=0$.
For $\alpha\neq 0$, at least $N> 2\pi/|{\rm Arg} \, t_\pm|$ orders are required to see the alternating nature of the series.

Even if $E_0(\lambda,\alpha)$ is Borel resummable for $|\alpha|<1$ in SPT, EPT can be used to greatly improve the numerical results at strong coupling.
Indeed, we can define a potential  $\hat V=V_0 + \lambda V_1$ with \be
V_0= \frac{1}{2}x^2 + \frac{\lambda}{2}x^4 \,, \quad \quad   V_1= \alpha  \frac{\sqrt{\lambda}}{\lambda_0} x^3 \,,
\label{eq:V-TiltedHatAnh}
\ee
so that the original one is recovered for $\lambda_0=\lambda$.

The first terms in the perturbative SPT and EPT expansions read
\bea
E_0  & \! = \!& \frac{1}{2} + \frac{3-11 \alpha ^2}8 \lambda
   - \frac{21-342 \alpha^2 +465 \alpha ^4}{32} \lambda^2
  + \frac{333-11827 \alpha^2+45507 \alpha ^4
     -39709
   \alpha ^6}{128} \lambda^3 +\ldots 
   \nn \\
\hat E_0 & \! = \! & \frac{1}{2} +\frac{3}{8} \lambda -\frac{21}{32} \lambda^2  + \Big(\frac{333}{128}- \frac{11 \alpha^2}{8\lambda_0^2} \Big)  \lambda^3  +\ldots\,,
\label{E0Ehat0}
  \eea
which shows how EPT rearranges all the $\alpha$-dependent terms in the perturbative expansion. For instance, the one-loop $\alpha^2$-dependent term in SPT appears at three loops 
in EPT. 

As we discussed, $V_1$ modifies the overall normalization of the large-order coefficients $\hat E_{0,n}(\lambda_0)$ with respect to the ones of the anharmonic oscillator $E_{0,n}(\alpha=0)$ without altering their leading large-order $n$-dependence. 
The normalization at leading order is given by the exponential  of the integral of $V_1$ evaluated at the nearest complex saddles
\be
\exp\Big(-\alpha \frac{\lambda^{3/2}}{\lambda_0} \int_{-\infty}^{\infty} \! d\tau \, z^3_\pm[\tau] \Big)= e^{\pm i \pi \alpha/2\lambda_0}\,.
\label{eq:ex0}
\ee
In analogy to the one-dimensional case outlined at the end of sect. 2, we expect for $n\gg 1$ and $n\gg 1/\lambda_0^2$,
\be
\hat E_{0,n}\Big(\frac{\lambda_0}{\alpha}\Big) =  E_{0,n}(\alpha=0)
\left[\cos\Big(\frac{\pi\alpha}{2\lambda_0}\Big)+{\cal O}\Big(\frac 1n\Big)\right]\,,
\label{eq:ex1}
\ee
where
\begin{equation}
E_{0,n}(\alpha=0)=-\frac{\sqrt6}{\pi^{3/2}} \left(-\frac{3}{2}\right)^n \Gamma\left(n+\frac12\right)\left[1 +
 {\cal O}\Big(\frac 1n\Big)\right] \,.
\end{equation}
In particular, eq.~(\ref{eq:ex1}) implies that the leading singularity of the Borel function is located at $t=-2/3$ as in the case of the anharmonic oscillator with $\alpha=0$. 
This is numerically confirmed by the associated Borel-Pad\'e approximants (see fig.~\ref{fig:BorelPlane}).
It is useful to compare the efficiency of  SPT and EPT  as a function of the number of terms $N$ that are kept in the series expansion.
These are reported  in fig.\ref{fig:TA-accuracy} for weak and strong coupling values $\lambda=1/20$,  $\lambda=5$, respectively, where
$\Delta E_0$ refers to the discrepancy with respect to $E_0^{{\rm RR}}$.  

In agreement with expectation, at sufficiently weak coupling SPT performs better than EPT.
The situation is drastically different at strong coupling, where SPT is essentially inaccurate for any $N$ reported in fig.\ref{fig:TA-accuracy}, while EPT has an accuracy that increases
 with the order.

At fixed number of perturbative terms, EPT works 
at its best for coupling constants $\lambda\sim {\cal O}(1)$. Like SPT, as $\lambda$ increases 
the integral in eq.~(\ref{BorelLeRoy}) is dominated by larger values of $t$ (this is best seen by rescaling $t\rightarrow t/\lambda$) and hence more and more terms of the perturbative expansion
are needed to approximate the Borel function. On the other hand, in analogy to the one-dimensional case (\ref{BorelIntExp0M}), the Borel function in EPT contains an additional exponential term coming from $\exp(-\int V_1)$. When $\lambda\ll 1$
the accuracy drops because the  coefficients $\hat E_n(\lambda_0)$ are very large before reaching the regime when eq.~(\ref{eq:ex1}) applies.

\subsection{Symmetric Double-well}

For $|\alpha|=1$ the potential (\ref{eq:V-TiltedAnh}) turns into a symmetric double-well, with two degenerate minima.
This is the prototypical model where real instantons are known to occur and the perturbative 
expansion around any of the two minima are known to be not Borel resummable. SPT requires the 
addition of an infinite number of real instantons to fix the ambiguities related to lateral Borel resummations and to reproduce the full result, see e.g. ref.\cite{Jentschura:2001kc} for a numerical study. 

\begin{table}[t]
\centering
{\renewcommand{\arraystretch}{1.2}%
\begin{tabular}{| c| c | c | c|c|c|c|}
 \hline
  $k$ &  $x=1/8$ & $x=1/4$ & $x=1/2$   &  $x=1$  & $x=2$ & $\Delta E_n/E_n$  \\ \hline
\hline
\multicolumn{7}{|c|}{Anharmonic}   \\ 
\hline 
 0   &   $7\cdot 10^{-24}$   &   $4\cdot 10^{-23}$  & $2\cdot 10^{-22}$   &  $10^{-21}$ & $8\cdot 10^{-20}$ & $3\cdot 10^{-33}$ \\
 1   &   $3\cdot 10^{-15}$  &   $2\cdot 10^{-14}$   & $7\cdot 10^{-14}$  &   $5\cdot 10^{-13}$   & $3\cdot 10^{-11}$ &$4\cdot 10^{-30}$ \\
 2   &   $2\cdot 10^{-8}$  &   $10^{-7}$   &   $4\cdot 10^{-6}$    & $6\cdot 10^{-7}$ & $7\cdot 10^{-6}$ &$2\cdot 10^{-27}$ \\ \hline
\multicolumn{7}{|c|}{Symmetric double well} \\ \hline
 0   &   $5\cdot 10^{-27}$  &   $3\cdot 10^{-26}$  & $10^{-25}$   &  $2\cdot 10^{-25}$ & $3\cdot 10^{-24}$ &$6\cdot 10^{-34}$  \\ 
 1   &   $4\cdot 10^{-18}$  &   $2\cdot 10^{-17}$   & $10^{-16}$  &   $6\cdot 10^{-16}$   & $2\cdot 10^{-14}$ &$10^{-30}$ \\ 
 2   &   $5\cdot 10^{-11}$  &   $3\cdot 10^{-10}$   &   $3\cdot 10^{-9}$    & $2\cdot 10^{-9}$ & 
 $10^{-8}$ &$5\cdot 10^{-27}$ \\ \hline
\end{tabular}
}
\caption{\small  Relative errors of the ratio of wave functions (\ref{eq:WF}) and energies of the first three levels of the anharmonic and symmetric double well at $\lambda=1$, 
evaluated at different points $x$ using EPT with $N=200$, and RR methods. In the anharmonic case EPT coincides with SPT. We have taken $x_0=1/16$ in eq.~(\ref{eq:WFx0}).}
\label{table:WF}
\end{table}

Shifting the $x$ coordinate so that $x=0$ is the maximum of the potential
\begin{equation} \label{eq:sdw0}
V(x;\lambda)=\frac{\lambda}{2}\left( x^2-\frac{1}{4\lambda}\right)^2\,,
\end{equation}
we can perform an EPT by considering the auxiliary potential 
\be
V_0 =\frac{1}{32\lambda} + \frac{\lambda_0}2 x^2+\frac{\lambda}2 x^4\,,
\qquad V_1=- \Big(1+\frac{1}{2\lambda_0}\Big) \frac{x^2}2 \,,
\label{eq:sdw1}
\ee
which has as effective couplings $\lambda/\lambda_0^{3/2}$ and $\lambda/\lambda_0 (1+1/(2\lambda_0))$.
This choice of EPT, where the minimum of $V_0$ is half way between the two minima of the double well, 
is such that the numerical 
Borel-Pad\'e resummation is able to reconstruct the non-perturbative splitting between the first two levels  at moderately small couplings, with a few hundred orders of perturbation theory. 
However, at fixed order $N$ of perturbation theory and for very small couplings, the true vacua depart further and further from the minimum of $V_0$ and the corresponding EPT becomes even worse than the naive truncated series.  In this regime a better choice would be to take the minimum of $V_0$ 
 close to one of the true minima of the double well (although resolving
 non-perturbative effects in this regime becomes harder and, as expected, 
 more terms of the perturbative expansion are required).
 
 We start by considering the ground state energy $E_0(\lambda)$. The large order behavior of
 the series coefficients $\hat E_{0,n}(\lambda_0)$ in EPT for $n\gg 1$ and $n\gg 1/\lambda_0^2$
are given by 
\be
\hat E_{0,n}(\lambda_0) = \lambda_0^{\frac12-\frac32 n} \, e^{-\sqrt{\lambda_0}\big(1+\frac{1}{2\lambda_0}\big)} E_{0,n}(\alpha=0)\left[1+{\cal O}\Big(\frac 1n\Big)\right]\,.
\label{eq:ex1a}
\ee
As before, the exponential $\lambda_0$-dependent factor is obtained by evaluating the potential $V_1$, the second term in square brackets in eq.~(\ref{eq:sdw1}), at the leading complex instanton solutions $z_\pm$.
The prefactor $ \lambda_0^{\frac12-\frac32 n}$  is instead due to the $\lambda_0$ dependence
of the quadratic term in $V_0$.

By taking $N=200,\lambda=\lambda_0 = 1/32$, we get $\Delta E_0/E_0\approx 2\cdot 10^{-5}$ 
and $\Delta E_1/E_1\approx 2\cdot 10^{-11}$.
These accuracies are already several orders of magnitude smaller than the leading order one-instanton contribution 
\be
E_0^{\rm inst} \approx \frac{2}{\sqrt{\pi \lambda}} e^{-\frac{1}{6\lambda}}\,,
\label{eq:sdw2}
\ee
which amounts to $\approx$$\,0.031$, or from the whole instanton contribution computed as the energy split between the ground state and the first excited level, which amounts to $\approx$$\,0.024$. 
For larger values of the coupling $\lambda$ the accuracy of EPT improves very quickly. For instance, already at $\lambda=\lambda_0 = 1/25$, keeping $N=200$ as before,
$\Delta E_0/E_0\approx 10^{-8}$ and $\Delta E_1/E_1\approx 4\cdot 10^{-14}$,
way smaller than the leading one-instanton contribution $\approx$$\,0.087$ or of the whole instanton contribution, computed as above and $\approx$$\,0.061$. 
For $\lambda\geq 1$ SPT breaks down: one would need to resum the whole transseries
 given by the multi-instantons and their saddle-point expansions, which is very challenging.
On the other hand EPT works very efficiently in this regime. 
At fixed order $N$, the error decreases as $\lambda$ increases
 up to some value, beyond which the error slowly increases again.  
 There is no need to consider too large values of $N$ to get a reasonable accuracy, in particular in the strong coupling regime  $\lambda\sim 1$.
 
For instance, at $\lambda=1$ and with $N=2(4)$ orders, we get $\Delta E_0/E_0 \simeq 3\%$($0.5\%$) 
by using the conformal mapping method \cite{loeffel,LeGuillou:1979ixc} with coupling
\be
w(\lambda) = \frac{\sqrt{1+3\lambda/2}-1}{\sqrt{1+3\lambda/2}+1}\,,
\ee
and Borel resumming the new series.
 \begin{figure}[t!]
\centering
\includegraphics[scale=.35]{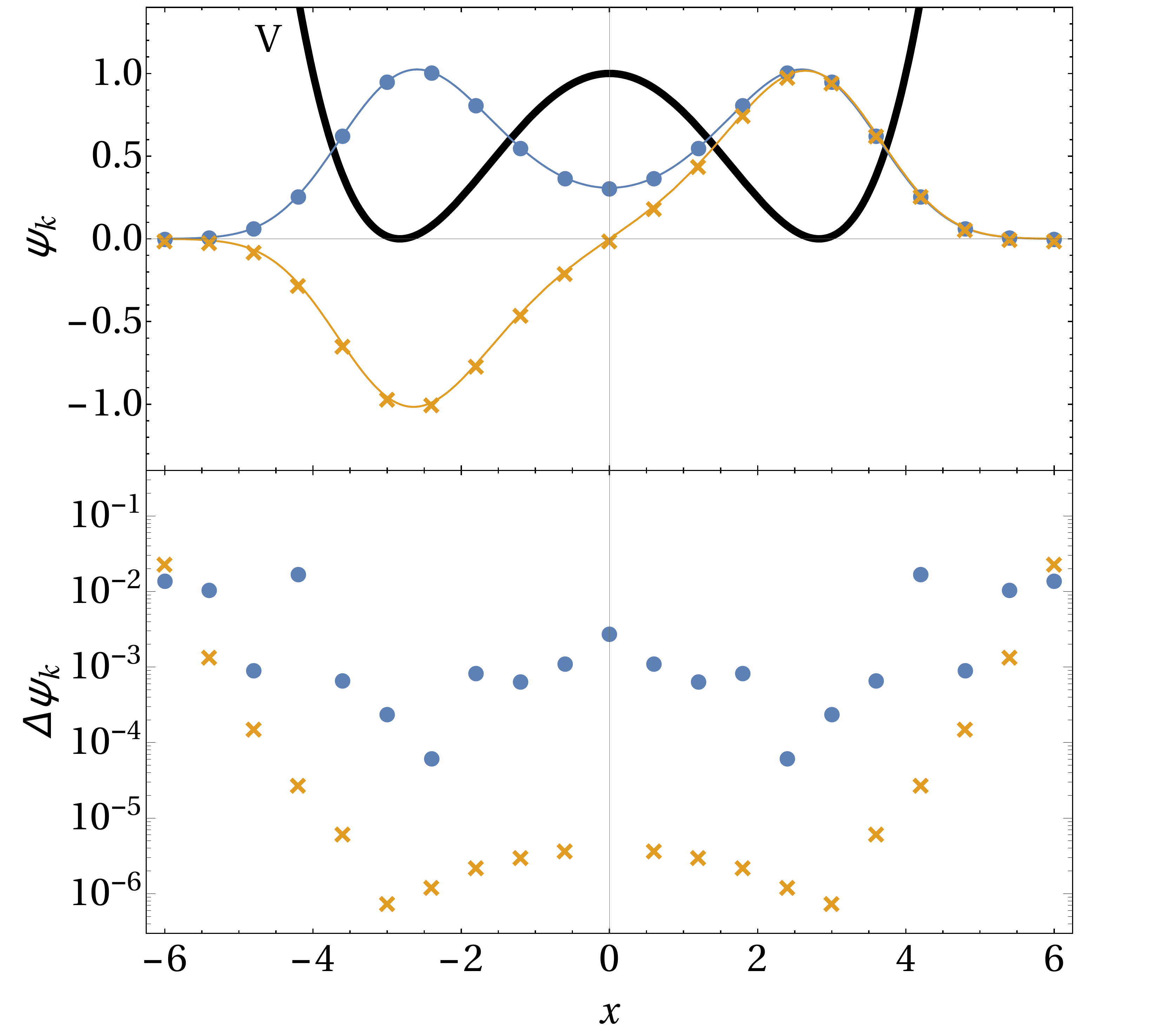}\qquad
\caption{\small
Comparison of the wavefunctions (normalized as in eq.~(\ref{eq:WFx0}) at $x_0=2\sqrt2$) 
for the first two levels of the symmetric double-well potential with $\lambda=1/32$ . 
Those computed in EPT with $N=200$ are
indicated by blue bullets (ground state) and orange crosses (first state) while the ones computed 
using the RR method are indicated by solid curves. 
}
 \label{fig:symdw}
\end{figure}

As far as we know, the convergence of series related to observables other than energy eigenvalues have been poorly studied.
This has motivated us to analyze  the series associated to the wave functions $\phi_k(x;\lambda)$.
We report in table \ref{table:WF} the values of $\Delta \psi_k(x;\lambda)$ for the first three levels of the anharmonic oscillator, $\alpha=0$ in eq.~(\ref{eq:V-TiltedAnh}),
and the symmetric double well (\ref{eq:sdw0}), for some values of $x$ at $\lambda=1$. 
Given the exponential decay of the wave function, larger values of $x$ are subject to an increasing numerical uncertainty
and are not reported. The decrease in accuracy as the level number $k$ increases is also expected, since both the RR methods and SPT/EPT require
more and more precision. 
In all the cases considered the Borel-Pad\'e approximants are free of poles in the real positive $t$ axis. 
The results clearly indicate that EPT captures the full answer. 
For illustration in fig.~\ref{fig:symdw} we plot $\psi_{0,1}$ and $\Delta \psi_{0,1}$ at $\lambda=1/32$. 

\subsection{Supersymmetric Double Well}

We now turn to the notable tilted double-well potential
 \be
V(x;\lambda) = \frac{\lambda}2\Big(x^2-\frac{1}{4\lambda}\Big)^2 + \sqrt{\lambda} x \,.
\label{eq:sdw4}
\ee
This is the exact quantum potential that one obtains from the supersymmetric version of the double well when the fermionic variables are integrated out.
As it is well known, the ground state energy $E_0=0$ to all orders in SPT due to supersymmetry. At the non-perturbative level, however, $E_0\neq 0$
because supersymmetry is dynamically broken \cite{Witten:1981nf}. Due to the absence of perturbative contributions in SPT, the supersymmetric double-well
is the ideal system where to test EPT.  It is also one of the simplest system where 
a perturbative expansion is Borel resummable
(being identically zero), but the sum does not converge to the exact value. 

Different authors have invoked complex instantons to reproduce $E_0$ \cite{Balitsky:1985in,Behtash:2015zha}. 
Their argument is essentially based on the observation that the entire quantum tilted potential
(\ref{eq:sdw4}) does not admit other real saddles that can contribute to the ground state energy.
Note however that the perturbation theory in $\lambda$ 
corresponds to the expansion around the saddle points of the classical action.
Since the tilt in eq.~(\ref{eq:sdw4}) is quantum in nature, 
the saddle-points of this system are the same as the ones of the symmetric double-well. In particular, real instantons occur, meaning that the path integral is on a Stokes line. 
The instanton contributions to $E_0$ have been extensively studied in ref.\cite{Jentschura:2004jg}, where the first nine terms of the perturbative series around the 1-instanton saddle have been computed using a generalized Bohr-Sommerfeld quantization formula \cite{ZinnJustin:1983nr}. As expected, this expansion agrees very well with the
numerical calculation at small coupling, while it breaks down when $\lambda$ approaches one
since the perturbative expansion of more and more instantons need to be properly included.

Note that the expansion around the saddle-points of the full action corresponds to treat the whole potential (\ref{eq:sdw4}) as classical. This means that
the coefficient of the linear tilt is rescaled by a factor $\lambda_0/\lambda$ 
to satisfy eq.~(\ref{Ex0}).
The resulting potential, which leads to an alternative perturbative expansion 
in $\lambda$ (APT), reads
\begin{equation} \label{eq:VAPT}
V_{\rm APT}=\frac{\lambda}2\Big(x^2-\frac{1}{4\lambda}\Big)^2 + \frac{\lambda_0}{\sqrt{\lambda}} x \,.
\end{equation} 
The original result is recovered by setting $\lambda=\lambda_0$.
This expansion for the ground state energy is no longer supersymmetric, but according to the discussion in  subsec. 3.2  is Borel resummable to the exact result.
\begin{figure}[t!]
\centering
\includegraphics[scale=.35]{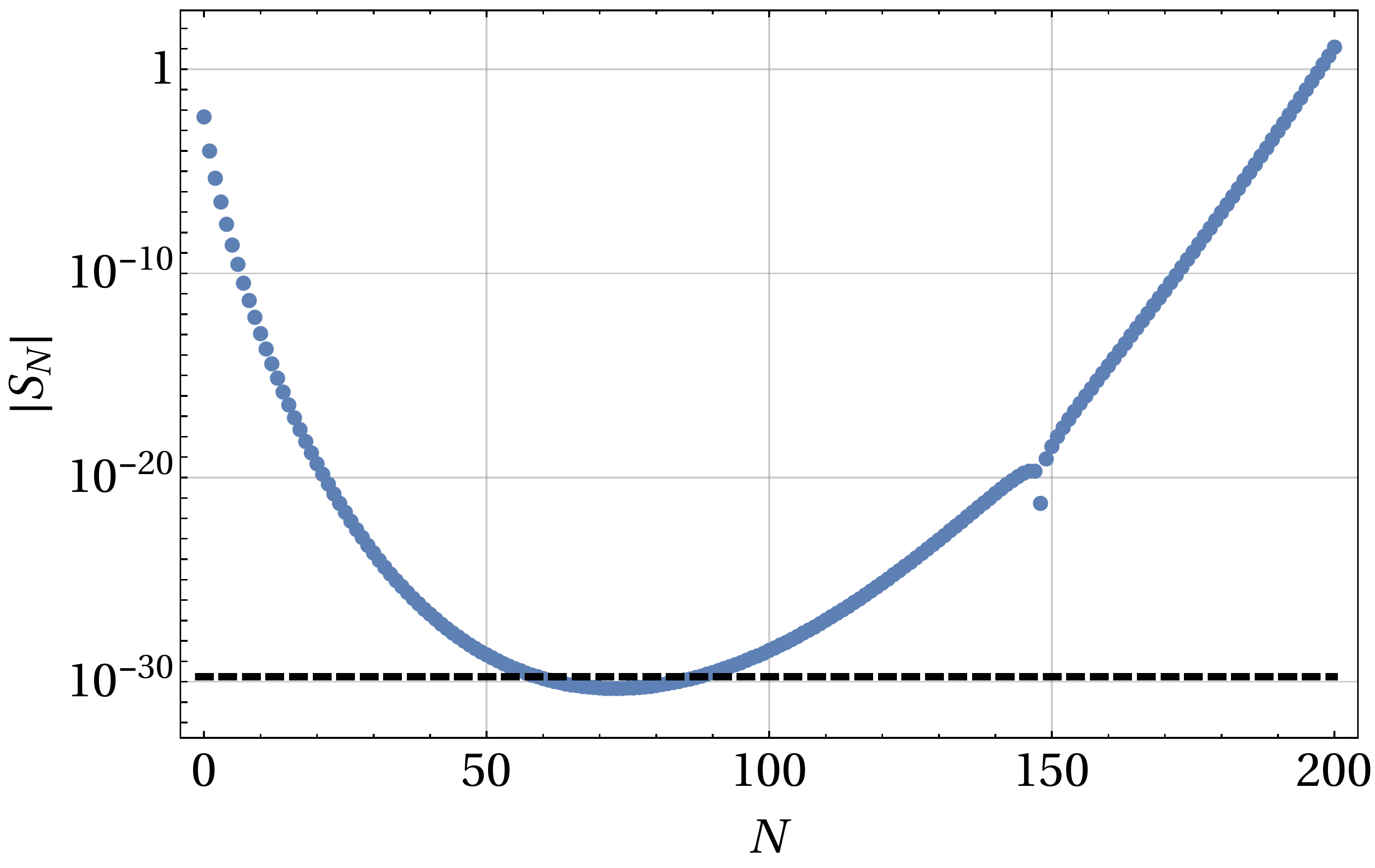}\qquad
\caption{\small 
The partial sum (\ref{eq:SN}) of APT for the potential (\ref{eq:VAPT}) 
as a function of $N$ for $\lambda = 1/200$.
The black dashed line corresponds to the exact ground state energy as computed using RR methods.}
 \label{fig:SUSYSDW-v2}
\end{figure}
We show in fig.~\ref{fig:SUSYSDW-v2} the partial sums 
\be \label{eq:SN}
S_N=\sum_{k=1}^N c_k \lambda^k
\ee
of the coefficients of APT as a function of $N$ for $\lambda = 1/200$. The dashed line represents the exact ground state energy. While each term in perturbation theory is non-vanishing, cancellations make the size of the truncated sum to decrease until 
$N\approx N_{\rm Best}=74$, where it approaches the expected
non-perturbative answer. For larger values of $N$ the series starts diverging, though  
Borel resummation reconstructs the right value with a precision of ${\cal O}(10^{-3})$ using $N=200$ terms. The feature appearing around $N=150$ is due to the change of sign
of the truncated sum. Indeed the sign of the coefficients $c_k$ oscillates 
with a long period ${\cal O}(150)$ since at weak coupling the tilt of the double-well potential is small and complex singularities of the Borel plane 
are close to the real axis (see fig.~\ref{fig:BorelPlane}).

Analogously to the previous cases we can also introduce an EPT 
for which all the observable are Borel resummable. 
For this purpose we consider the auxiliary potential 
 \be
V_0= \frac{1}{32\lambda} + \frac{\lambda_0}2 x^2+\frac{\lambda}2 x^4\,, \quad \quad V_1= \frac{x}{\sqrt{\lambda}}  - \left (1+\frac{1}{2\lambda_0}\right ) \frac{x^2}2 \,,
\label{eq:sdw4a}
\ee
where $V_1$ includes the quantum tilt $x$ and the quadratic $x^2$ term necessary to recover the original potential. The specific decomposition (\ref{eq:sdw4a}) turns out to be numerically convenient for moderately small and large couplings.
 \begin{figure}[t!]
\centering
\includegraphics[scale=.35]{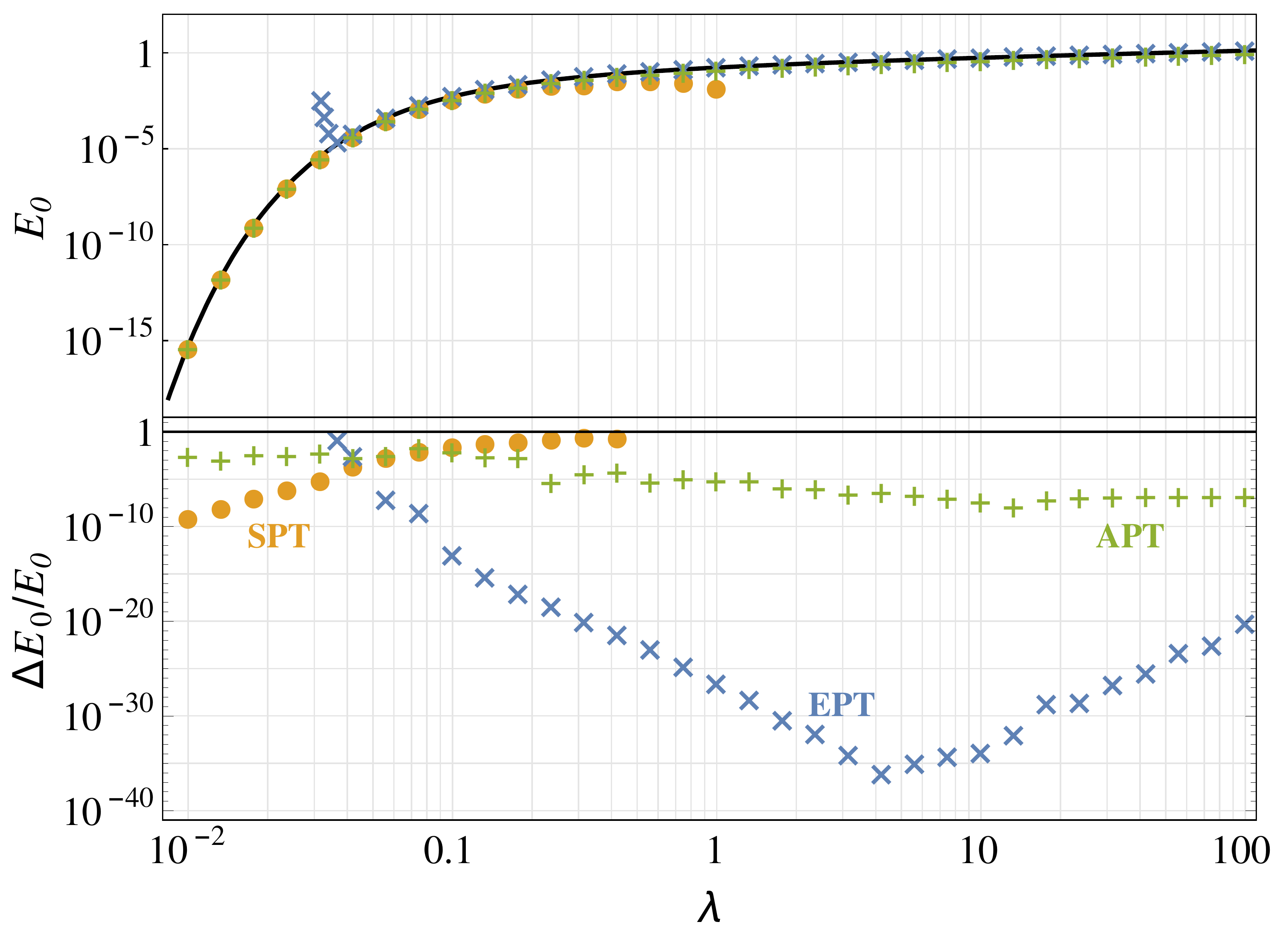}\qquad
\caption{\small The ground state energy (top) and the relative error (bottom) as a function of the coupling $\lambda$ for the supersymmetric double well (\ref{eq:sdw4}). 
The blue [green] crosses  refer to EPT [APT] from the potential (\ref{eq:sdw4a}) 
[(\ref{eq:VAPT})] with $N=200$ series coefficients, the orange dots to SPT of ref.\cite{Jentschura:2004jg}, with a truncated expansion up to the ninth order around the leading instanton. The black line corresponds to the exact result, computed by means of a Rayleigh-Ritz method.}
 \label{fig:SUSYSDW}
\end{figure}

In fig.~\ref{fig:SUSYSDW} we show a comparison between the various perturbative
estimates of $E_0$ and the numerical RR one as a function of the coupling constant. 
As expected, at small coupling SPT provides the best estimate.
In fact it encodes analytically the leading instanton effect 
providing the asymptotic value of $E_0$ at $\lambda\to 0$. 
However, already at moderately small couplings both APT and EPT are able to
resolve the leading instanton effects with a good accuracy.
At moderate and strong coupling the instanton computation quickly breaks down, while both APT and EPT work extremely well. 
In particular the accuracy of EPT strongly increases with $\lambda$ up to $\lambda\sim 4$.
For larger values of the coupling the accuracy drops, but it remains remarkable. For $\lambda \sim 10^2$ we have an accuracy $\sim 10^{-20}$ with 200 orders of perturbation theory. It is amazing how a perturbative computation can work so efficiently at strong coupling!

\subsection{False Vacuum}
\begin{figure}[t]
\centering
\includegraphics[scale=.35]{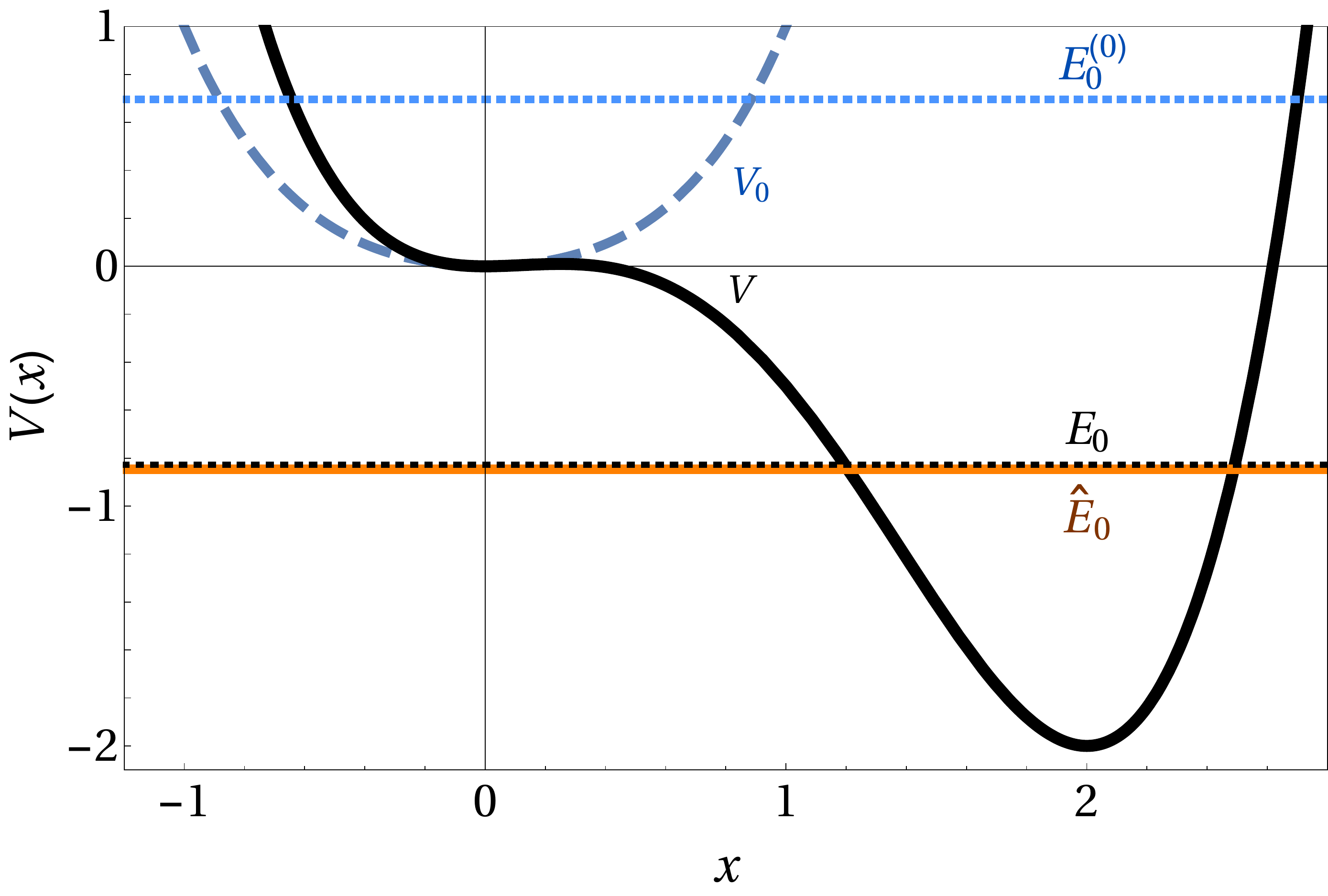}\qquad
\caption{\small Plots of the potential (\ref{eq:Vfv})  (solid black curve) and the associated potential $V_0$ in eq.~(\ref{eq:V0fv})  (dashed curve) with $\lambda=1$.
The dashed blue line corresponds to the ground state energy associated to $V_0$, the solid red and the dashed black ones are the ground state energy of the potential (\ref{eq:Vfv})
obtained from EPT and RR methods, respectively.}
\label{fig:falsevacuum}
\end{figure}

We now consider the potential (\ref{eq:V-TiltedAnh}) with $|\alpha|>1$. In this case $x=0$ is no longer the absolute minimum, and we are effectively expanding around a false vacuum.
Clearly SPT is non-Borel resummable in this case, given the presence of other 
real instantons. The perturbative expansion around the
false vacuum does not contribute at all to $E_0$, which, as we saw,
is entirely reconstructed by the expansion around the true vacuum. 
Still the EPT around the false vacuum 
defined by the potential (\ref{eq:V-TiltedHatAnh})
is able to recover the (true) ground state energy.

In fig.~\ref{fig:falsevacuum}  we show the shape of the original potential $V$ for $\alpha=-3/2$
\be
V(x;\lambda)=\frac{1}{2}x^2 -\frac{3}{2} \sqrt{\lambda} x^3+ \frac{\lambda}{2}x^4 \,,
\label{eq:Vfv}
\ee
and the corresponding exact ground state energy $E_0\approx -0.828$ at $\lambda=1$.
We also show the potential $V_0$ used in EPT
\be
V_0=\frac{1}{2}x^2+\frac{\lambda}{2}x^4 \,,
\label{eq:V0fv}
\ee
with the would-be ground state energy $E_0^{(0)}\approx 0.696$.
Using EPT with $N=280$ orders such value moves to $\hat E_0 \approx -0.847$. 
Although the accuracy is not comparable to that obtained in the previous cases, it is remarkable that one is able to compute the energy of the true vacuum starting
from a perturbative expansion around a false vacuum.

\subsection{Degenerate Saddle Points: Pure Anharmonic Oscillators}

In this subsection we discuss how to use EPT to address the infinitely coupled
systems described by potentials with degenerate saddle points. Consider for example 
the pure anharmonic oscillators with potentials of the form
\be
V(x) = 2^\ell x^{2\ell} \,,\qquad \ell\in \mathbb{N}^+\,.
\label{pureHan}
\ee
The factor $2^\ell$ is such that, modulo a trivial rescaling, the Hamiltonian is of the form $p^2+x^{2\ell}$ which is 
the conventional normalizaton used in the literature for this class of models.
Pure anharmonic oscillators are intrinsically strongly coupled and can be obtained as the $\lambda\rightarrow \infty$ limit of their corresponding ordinary massive anharmonic oscillators
after the rescaling $x\rightarrow \lambda^{-1/2(1+\ell)}x$. The potentials (\ref{pureHan}) are convex with a degenerate minimum at $x=0$. 
In the absence of a quadratic mass term, perturbation theory cannot be used. The energy eigenvalues $E_k^{(2\ell)}$ of these systems have instead been studied using 
Rayleigh-Ritz methods (see e.g. ref.\cite{Taseli}), Thermodynamic Bethe Ansatz  (TBA) \cite{Dorey:1998pt} or a Wentzel Kramers Brillouin (WKB) approximation \cite{BPV} (see also ref.\cite{Grassi:2014cla} for a more modern perspective). In the WKB approximation one considers a series expansion in $1/k$, $k$ being the quantum number level.
It was found in refs.\cite{BPV,Grassi:2014cla} that the asymptotic series of the WKB expansion, where classical real trajectories in phase space are considered, 
does not reproduce the correct result. A better accuracy is achieved by adding in the WKB quantization formula the contribution of complex trajectories in phase space.
However, there are an infinite number of them and a parametrically high accuracy could be obtained only by resumming all the infinite complex trajectories.
 
In terms of Lefschetz thimbles, the potentials (\ref{pureHan}) have a degenerate saddle for which our considerations do not apply. 
A possibility is to add a mass term $\epsilon x^2$ to eq.~(\ref{pureHan}), compute the energy levels $E_k^{(2\ell)}(\epsilon)$ and extrapolate
the result for $\epsilon\rightarrow 0$. By choosing $\epsilon>0$ we are guaranteed that $E_k^{(2\ell)}(\epsilon)$ are Borel resummable for any $\epsilon$
and no (real or complex) non-perturbative contributions are expected. We have verified this expectation by computing the ground state energy $E_0^{(4)}(\epsilon)$ for the pure quartic oscillator 
for smaller and smaller values of $\epsilon$ (using the Pad\'e-Borel method) and have found that the extrapolated value $E_0^{(4)}$ converges to the exact value.

The same result can be found with much greater accuracy and efficiency using EPT
without taking any extrapolation. Consider the auxiliary family of potentials defined as
\be
V_0 =  \lambda^{\ell-1} 2^\ell x^{2\ell}+\sum_{j=1}^{\ell-1} c_j \lambda^{j-1} x^{2j}\,, \qquad 
V_1 = - \frac{1}{\lambda_0}\sum_{j=1}^{\ell-1} c_j \lambda^{j-1} x^{2j}\,,
\label{pureHan4}
\ee
such that at $\lambda=\lambda_0=1$ the potential in eq.~(\ref{pureHan}) is recovered. 
By a proper choice of the $\ell-1$ coefficients $c_j$, $V_0$ has a unique non-degenerate minimum at $x=0$ and perturbation theory is well-defined. 

For the pure quartic case $\ell=2$, by choosing $c_1=2$ in eq.~(\ref{pureHan4}) and by  using only the first ten orders of  EPT we get $E_0^{(4)} \simeq 1.060362$, which is more accurate than the value in tab. 2 of ref.\cite{Grassi:2014cla}, obtained using 320 orders in the WKB expansion with the inclusion of the leading complex saddles.
The accuracy is easily improved using more coefficients of the perturbative expansion.  
\begin{table}[t]
\centering
{\renewcommand{\arraystretch}{1.4}%
\begin{tabular}{| c| c | c | c|c|c|}
\hline
 $k$    &    
 $E_k^{(4)}$      &    $\Delta E_k^{(4)}/E_k^{(4)}$     &   $E_k^{(6)}$      &    $\Delta E_k^{(6)}/E_k^{(6)}$  \\
\hline
 0    &    1.0603620904   &     $3 \cdot10^{-45}$   & 0.5724012268    &     $2 \cdot 10^{-19}$  \\
 1    &    3.7996730298   &     $2 \cdot 10^{-44}$    & 2.1692993557    &     $3 \cdot 10^{-19}$   \\ 
 2    &    7.4556979379   &     $9 \cdot 10^{-37}$    & 4.5365422804    &     $9 \cdot 10^{-17}$   \\ 
 3    &    11.6447455113   &     $4 \cdot10^{-36}$   & 7.4675848174    &     $7 \cdot10^{-16}$   \\
 4    &    16.2618260188    &     $4 \cdot 10^{-36}$   & 10.8570827110    &     $2 \cdot 10^{-16}$ \\ \hline
\end{tabular} }
\caption{\small Energy eigenvalues $E_k^{(2\ell)}$ and the corresponding accuracies $\Delta E_k^{(2\ell)}/E_k^{(2\ell)}$ of the first five levels of the pure anharmonic $x^4$ and $x^6$ potentials,  computed using EPT with $N=200$. Only the first ten digits after the comma are shown (no rounding on the last digit).}
\label{table:Pure}
\end{table}
We have computed in this way the $E_k^{(2\ell)}$ for different values of $\ell$ and $k$, as well as the associated wave-functions $\psi_k^{(2\ell)}(x)$ for some values of $x$.
In all cases we found excellent agreement between our results and those obtained with RR methods.

For illustration we report in tab.~\ref{table:Pure} the accuracies for the energy levels of the first five states of the pure $x^4$ and $x^6$ oscillators 
computed comparing EPT to the results from RR methods.
We used $N=200$ orders of perturbation theory and in eq.~(\ref{pureHan4}) 
we chose $c_1=2$ for $\ell=2$ and $c_1=4$, $c_2=2$ for $\ell=3$.
 Notice the accuracy of $E_0^{(4)}$ up to 45 digits!
At fixed $N$, similarly to the RR method, 
the accuracy decreases as the energy level and the power $\ell$ in eq.~(\ref{pureHan}) 
increase (in contrast to the WKB method where the opposite occurs)
All the energy eigenvalues reported in tab. \ref{table:Pure} are in agreement with those reported in tab. 1 of ref.\cite{Taseli}, tab. I and II of ref.\cite{Dorey:1998pt}
and tab. 2 of ref.\cite{Grassi:2014cla}, in all cases computed with less precision digits than our results.\footnote{Note however 
that the numerical computations based on the Rayleigh-Ritz 
methods remain superior for these simple potentials.} 
The accuracy of our results sensibly depend on the choice of the coefficients
$c_j$ in eq.~(\ref{pureHan4}). We have not performed a systematic search of the optimal choice that minimizes the errors, so it is well possible that at a fixed order $N$ a higher accuracy than that reported in tab.~\ref{table:Pure} can be achieved. 

\section{Conclusions and Future Perspectives in QFT}  \label{sec:concl}

In this paper we have studied one-dimensional QM systems with bound-state potentials and discrete spectra. We characterized some of the conditions for the Borel summability of perturbation theory
in QM. In particular when the potential admits only one critical point (minimum), we have shown that the loopwise expansion in the Euclidean path integral is Borel resummable and reproduces the full answer. Several known results in the literature about the Borel summability of certain QM systems, such as the quartic anharmonic oscillator  \cite{Loeffel:1970fe,Simon:1970mc}, are rederived and generalized in this new perspective.

We also explained why EPT---the modified perturbative expansion introduced in 
ref.~\cite{shortpaper}---is able to extend the above result to generic bound-state potentials.
Remarkably, EPT encodes all the non-perturbative corrections of SPT, 
the standard semi-classical expansion including instantons, providing the full answer 
for any value of the coupling constant. In particular, EPT works at its best at strong coupling, where
the high accuracy obtained confirms its validity. 
All complications occurring in SPT, related to the need of a resurgence analysis to make sense of otherwise ambiguous results, or of a 
highly non-trivial Lefschetz thimble decomposition of the path integral, are bypassed when using EPT. 
These points have been illustrated in details with several examples. 

Our results can be extended in various directions.
It would be interesting to understand if and how they can be obtained directly in Minkowski space where, contrary to the Euclidean case, one would always expect an infinite number of saddles to contribute.  On the other hand the generalization to higher-dimensional QM systems with bound-state potentials should be straightforward.

The extension of our results to QFT is probably the most interesting.
Possible complications arise both from UV and IR effects. As it is well-known, in contrast to QM, in QFT it is not enough to properly define the path integral measure to get a finite theory.  The key point to address is understanding how the renormalization procedure needed to remove  UV divergencies affects 
the Lefschetz thimble approach to the path integral. 

The IR limit (i.e. infinite volume limit) is also more subtle than in QM because of possible phase transitions and spontaneous breaking of symmetries.
Since in this case observables $O(\lambda)$ may no longer be analytic for any value of the coupling $\lambda$, it is not clear what will be the fate of the properties of perturbation theory and of the thimble decomposition.

Aside from the subtleties mentioned above, a naive extrapolation of our anharmonic potentials to scalar 
QFT  would lead to the expectation that perturbation theory for 
superrenormalizable potentials with a single critical point 
are Borel resummable to the exact result.
 The proof of the Borel summability for the particular cases of the $\phi^4$ theories with a positive squared mass term in $d=2$ and $d=3$ \cite{Eckmann,Magnen} seems to be compatible with this conjecture. 
The $\phi^4$ theory in $d=2$ seems the ideal laboratory to start exploring our ideas in the QFT framework. Indeed this is one of the simplest non-integrable QFT 
where UV divergencies can be removed by just normal ordering. 
In addition, in the infinite volume limit, it undergoes a second-order phase transition to a ${\bf Z}_2$-breaking phase in $d=2$  \cite{Chang:1976ek}.  
We hope to come back to the analysis of this model in the future.

On a more general perspective, by a proper analytic continuation in the space-time dimension, one might hope to put on firmer grounds the conjectured Borel summability of the $\epsilon$-expansion in the 
$\phi^4$ theory \cite{Brezin:1976vw}. In QFT we will not have access to many terms in the perturbative expansion, and of course we cannot expect the degree of accuracy 
that is possible in QM. Nevertheless, computations of critical exponents in the three-dimensional Ising and vector $O(N)$ models
have shown that  an accuracy at the per mille level can be achieved at strong coupling $\epsilon=1$ by resumming (using Borel-Pad\'e approximants, conformal mapping or other methods) the first few loops in perturbation theory \cite{Pelissetto:2000ek}. More ambitious goals would be to extend our methods to QFT that are {\it not} Borel resummable, like gauge theories in $d=4$. Any progress in this direction would be of great interest.

We think that the results of this paper have opened a new perturbative window on strongly coupled physics.

\section*{Acknowledgments}
We thank S. Cecotti, A. Grassi, J.~Elias-Mir\'o,  E.~Poppitz, T.~Sulejmanpasic and M. \"Unsal for discussions, and
G. Martinelli for having pointed out a mistake in the first version of eq.(\ref{eq:simpleBorel}).
This work is supported in part by the ERC Advanced Grant no.267985 (DaMeSyFla).

\appendix

\section{On the Finiteness of ${\cal B}_{-1/2}^{(\infty)} {\cal Z}$}

In this appendix we report some checks we performed about the finiteness of the continuum limit of ${\cal B}_{-1/2}^{(N)} {\cal Z}(\lambda t)$.
First of all, we consider the partition function of the harmonic oscillator.
We discretize the path integral by cutting-off the real Fourier modes coefficients $c_n$ of $x(\tau)$:
\be
x(\tau) = \sum_{k=0}^N c_k \eta_k \cos \frac{2\pi k \tau}{\beta} +  \sum_{k=1}^N c_{-k} \eta_k \sin \frac{2\pi k \tau}{\beta}\,, \quad \eta_k = \frac{\sqrt{2-\delta_{k,0}}}{\sqrt{\beta}} \,.
\ee
The discretized action reads
\begin{equation}
S^{(N)}[x]=\int_0^{\beta} d\tau\, \left[ \frac12 \dot x^2+\frac12 \omega^2 x^2\right]=
\frac12 \sum_{k=-N}^{N} \mu_k c_k^2 \,, \qquad \mu_k\equiv \omega^2+\frac{(2\pi k)^2}{\beta^2}\,.
\end{equation}
The path integral  measure reads
\begin{equation}
{\cal D}^{(N)}x(\tau)= \prod_{k=-N}^{N} N_k \, dc_k \,, \qquad N_k\equiv \left \{ 
\begin{array}{c c c }
\frac{|k|}{\beta} \sqrt{\frac{2\pi}{\lambda}} & \quad & k\neq 0 \\
\frac{1}{\beta\sqrt{2\pi \lambda}} & & k=0 
\end{array}
\right . \,.
\end{equation}
Introduce now a radial coordinate system defined as
\be
c_k = \sqrt{\frac{2}{\mu_k}} \rho \, \hat c_k\,, \quad \sum_{k=-N}^N \hat c_k^2 = 1
\ee
where $\rho$ is the radius and the $\hat c_k$'s encode the standard parametrization of the unit $2N$-sphere in terms of its $2N$-angles, whose explicit form will not be needed.
Using the expression (\ref{eq:ZBn2}) for the Borel-Le Roy function and the above results, we get
\begin{align}
{\cal B}_{-1/2}^{(N)}{\cal Z}_0(\lambda t) & =
\frac{ \sqrt{\lambda t}}{\beta} \prod_{k=1}^N \left[ \frac{2\pi k}{\beta}\right]^2 \partial_{\lambda t}^N 
\int_0^\infty \frac{d\rho}\rho\, \Omega_{2N+1} \left[ \prod_{k=-N}^N \frac{\rho}{\sqrt{\pi \mu_k}} \right]
 \delta(\rho^2-t\lambda)\nonumber  \\
&= \frac{1}{\Gamma\left (N+\frac12\right )} \sqrt{\lambda t}\,  \partial_{\lambda t}^N (\lambda t)^{N-\frac12} \frac{1}{\beta \omega} \prod_{k=1}^N \frac{1}{1+\left(\frac{\omega \beta}{2\pi k}\right)^2} \nonumber \\
&=\frac{1}{\Gamma\left(\frac12\right)}\frac{1}{\beta \omega} \prod_{k=1}^N \frac{1}{1+\left(\frac{\omega \beta}{2\pi k}\right)^2} \,,
\end{align}
where $\Omega_d=2\pi^{d/2}/\Gamma(d/2)$ is the area of the unit $(d-1)$-dimensional sphere and 
we used the relation
\be
\partial_{\lambda t}^N (\lambda t)^{N+p} = \frac{\Gamma(N+p+1)}{\Gamma(p+1)}  (\lambda t)^{p}\,,
\label{relation3}
\ee
valid for any value of $p$. Notice how taking $N$ derivatives with respect to $\lambda t$ gives rise to an $N$-dependent Gamma function that compensates the
one coming from the area of the unit-sphere. As expected, the dependence on $\lambda t$ disappears and the continuum limit gives the finite
answer
\begin{equation}
\lim_{N\to \infty} {\cal B}_{-1/2}^{(N)}{\cal Z}_0(\lambda t) \equiv \lim_{N\to \infty} \frac{{\cal Z}_0^{(N)}}{\Gamma(1/2)}=
\frac{ {\cal Z}_0}{\Gamma\left(\frac12\right)}\,, \quad\quad   {\cal Z}_0= \frac{1}{2\sinh\left(\frac{\omega \beta}{2}\right)}\,,
\label{ZNdef}
\end{equation}
which reproduces the known partition function $ {\cal Z}_0$ of the harmonic oscillator after the integral over $t$ is performed.

An exact computation of ${\cal B}_{-1/2}^{(N)} {\cal Z}(\lambda t)$ is clearly out of reach in interacting QM systems. 
Yet, we can show that ${\cal B}_{-1/2}^{(\infty)} {\cal Z}(\lambda t)$ is finite to all orders in perturbation theory for polynomial potentials.
It is useful to work out in detail the first order term of ${\cal B}_{-1/2}^{(N)} {\cal Z}(\lambda t)$ for the quartic anharmonic oscillator
$V(x)= \omega^2 x^2/2+x^4/4$. We have
\be
{\cal B}_{-1/2}^{(N)}{\cal Z}(\lambda t)  ={\cal Z}_0^{(N)}  \frac{\sqrt{\lambda t}}{\pi^{N+\frac 12}}  \partial_{\lambda t}^N 
\int_0^\infty d\rho\int d\Omega_{2N+1} \, \rho^{2N}
 \delta(\rho^2+\rho^4 \xi -t\lambda)\,,
 \ee
where ${\cal Z}_0^{(N)}$  is the discretized version of the harmonic oscillator partition function defined in eq.(\ref{ZNdef}) and
\be
\xi =\!\!\sum_{n,m,p,q=-N}^N \!\!\frac{\eta_n\eta_m\eta_p\eta_q}{\sqrt{\mu_n \mu_m\mu_p\mu_q}} \int_0^\beta \!\! d\tau  (\hat c_n X_n) (\hat c_m X_m) (\hat c_p X_p) (\hat c_q X_q)\,,\quad X_n\equiv \left \{ 
\begin{array}{c c }
 \cos \frac{2\pi n \tau}{\beta}\,,   & n\geq  0 \\
 \sin \frac{2\pi n \tau}{\beta}\,,   & n<0 
\end{array}
\right . \,.
\label{fdef}
\ee
At linear order in $\lambda t$, once we expand the argument of the delta function, we get
\be
{\cal B}_{-1/2}^{(N)}{\cal Z}(\lambda t)  ={\cal Z}_0^{(N)}  \frac{\sqrt{\lambda t}}{\pi^{N+\frac 12}}  \partial_{\lambda t}^N 
\int_0^\infty \!\!d\rho\!\int d\Omega_{2N+1} \rho^{2N-1} (1-2\rho^2 \xi)\delta[\rho - \sqrt{\lambda t} (1-\lambda t \xi/2)+\ldots ]+{\cal O}(\lambda t)^2 \,.
\label{B12quart}
 \ee
It is convenient to evaluate the integral over the angular variables $\hat c_n$ before the one in $d\tau$ appearing in eq.~(\ref{fdef}).
This is easily obtained by using the following identity, valid in cartesian coordinates in any number of dimensions $d$:
\be
\int \!d^dx \, x_n x_m x_p x_q f(x^2) = \frac{(\delta_{nm} \delta_{pq} +\delta_{np} \delta_{mq} +\delta_{nq} \delta_{mp})}{d (d+2)} \int \! d^dx \, x^4 f(x^2)\,, \quad x^2 = \sum_{k=1}^d x_k x_k\,,
\label{intxdx}
\ee
from which it immediately follows, taking $d=2N+1$,
\be
\int \!\! d\Omega_{2N+1} \hat c_n \hat c_m \hat c_p \hat c_q = \frac{(\delta_{nm} \delta_{pq} +\delta_{np} \delta_{mq} +\delta_{nq} \delta_{mp})}{(2N+1) (2N+3)}\Omega_{2N+1} = 
\frac{\pi^{N+\frac 12}}{2\Gamma(N+\frac 52)} (\delta_{nm} \delta_{pq} +\delta_{np} \delta_{mq} +\delta_{nq} \delta_{mp}).
\ee
The integral over $d\tau$ is straightforward and after a bit of algebra we get
\be
\int \!\! d\Omega_{2N+1} \xi =\frac{3\beta}2 \frac{\pi^{N+\frac 12}}{\Gamma(N+\frac 52)} ({\cal G}^{(N)})^2\,,\quad {\cal G}^{(N)} \equiv \frac1\beta \sum_{k=-N}^N \frac{1}{\mu_k} \,.
\label{Intfquart}
\ee
Plugging eq.~(\ref{Intfquart}) in eq.~(\ref{B12quart}) gives
\be\begin{split}
{\cal B}_{-1/2}^{(N)}{\cal Z}(\lambda t)  & = {\cal Z}_0^{(N)}  \Big(\frac{1}{\Gamma(1/2)}- \frac {3\beta}4  \sqrt{\lambda t} \partial_{\lambda t}^N (\lambda t)^{N+1/2}\frac{N+3/2}{\Gamma(N+\frac 52)} 
({\cal G}^{(N)})^2 +{\cal O}(\lambda t)^2 \Big) \\
& =  {\cal Z}_0^{(N)} \Big(\frac{1}{\Gamma(1/2)}- \frac {3\beta}4 ({\cal G}^{(N)})^2 \frac{\lambda t}{\Gamma(3/2)}  +{\cal O}(\lambda t)^2 \Big) \,.
\label{B12quart2}
\end{split} \ee
In the continuum limit  we have 
\be\begin{split}
\lim_{N\rightarrow \infty} {\cal B}_{-1/2}^{(N)}{\cal Z}(\lambda t)  =  {\cal Z}_0 \Big(\frac{1}{\Gamma(1/2)}- \frac 34\beta {\cal G}^2 \frac{\lambda t}{\Gamma(3/2)}  +{\cal O}(\lambda t)^2 \Big) \,,
\label{B12quart3}
\end{split} \ee
where 
\be
{\cal G} = \frac{1}{2\omega} \coth \frac{\beta \omega}2
\ee
is the particle propagator at $\tau=0$. After integrating over $t$, eq.~(\ref{B12quart3}) reproduces the first order perturbative correction to the partition function of the quartic anharmonic oscillator.

Finiteness of $ {\cal B}_{-1/2}^{(N)}{\cal Z}$ as $N\rightarrow \infty$ to all orders is easily shown. For simplicity, we just keep track of the factors of $N$, neglecting all other parameters. At order $(\lambda t)^k$, after expanding the argument of the delta function, we get 
\be
{\cal B}_{-1/2}^{(N)}{\cal Z}(\lambda t)|_{\lambda^k}  \propto {\cal Z}_0^{(N)}  \frac{\sqrt{\lambda t}}{\pi^{N+\frac 12}}  \partial_{\lambda t}^N  (\lambda t)^{N+k-1/2} N^{k} 
\!\int d\Omega_{2N+1}\xi^k  \,.
\label{B12all}
 \ee
The $4k$-generalization of eq.~(\ref{intxdx}) gives
\be
\int d\Omega_{2N+1}\xi^k \propto \frac{1}{N^{2k}} \Omega_{2N+1}\,.
\label{relation4}
\ee
Plugging eq.~(\ref{relation4}) in eq.~(\ref{B12all}) and using eq.~(\ref{relation3}) gives
\be
\lim_{N\rightarrow \infty}{\cal B}_{-1/2}^{(N)}{\cal Z}(\lambda t)|_{\lambda^k}  \propto \lim_{N\rightarrow \infty} {\cal Z}_0^{(N)}  \frac{(\lambda t)^k}{\pi^{N+\frac 12}} \frac{\Gamma(N+k+1/2)}{\Gamma(k+1/2)} \frac{\Omega_{2N+1}}{N^k} \propto {\cal Z}_0  \frac{(\lambda t)^k}{\Gamma(k+\frac 12)}   \,.
\label{B12allorders}
 \ee
Similarly, we can prove the finiteness of the continuum limit to all orders in perturbation theory for any other interaction term of the form $g\,x^{2p}$. Recall that
the loopwise parameter $\lambda$ corresponds to a coupling constant $g=\lambda^{p-1}$ and hence, for $p\neq 2$, the two are not identical. Taking that into account, the scaling in $N$
of ${\cal B}_{-1/2}^{(N)}{\cal Z}$ at order $g^k$ reads
\be
\lim_{N\rightarrow \infty}{\cal B}_{-1/2}^{(N)}{\cal Z}(\lambda t)|_{g^k}  \propto \lim_{N\rightarrow \infty} {\cal Z}_0^{(N)}  \frac{\sqrt{\lambda t}}{\pi^{N+\frac 12}}  \partial_{\lambda t}^N  (\lambda t)^{N+(p-1) k-\frac12} N^{k}  \frac{\Omega_{2N+1}}{N^{pk}} \propto \frac{g^k t^{k (p-1)}}{\Gamma\left[(p-1) k+\frac 12\right]} \,,
\label{B12allordersGen}
 \ee
and is finite. q.e.d.

Finiteness of ${\cal B}_{-1/2}^{(N)}{\cal Z}$ to all orders in perturbation theory for any polynomial potential term easily follows from the above results.

\end{document}